# Programmable lattices for non-Abelian topological photonics and braiding

Gyunghun Kim[1], Jensen Li[2], Xianji Piao[3§], Namkyoo Park[4†], and Sunkyu Yu[1*]

[1]Intelligent Wave Systems Laboratory, Department of Electrical and Computer Engineering, Seoul National University, Seoul 08826, Korea

[2]Department of Engineering, University of Exeter, EX4 4QF, United Kingdom

[3]Wave Engineering Laboratory, School of Electrical and Computer Engineering, University of Seoul, Seoul 02504, Korea

[4]Photonic Systems Laboratory, Department of Electrical and Computer Engineering, Seoul National University, Seoul 08826, Korea

E-mail address for correspondence: [§]piao@uos.ac.kr, [†]nkpark@snu.ac.kr, [*]sunkyu.yu@snu.ac.kr

## Abstract

Non-Abelian physics, originating from noncommutative sequences of operations, unveils novel topological degrees of freedom for advancing band theory and quantum computation. In photonics, significant efforts have been devoted to developing reconfigurable non-Abelian platforms, serving both as classical testbeds for non-Abelian quantum phenomena and as programmable systems that harness topological complexities. Here we establish topological spinor lattices for non-Abelian programmable photonics. We design a building block for reconfigurable unitary coupling between pseudospin resonances, achieving a universal set of rotation gates through coupling along the unit cell boundary. The lattice assembly of our building blocks enables the emulation of the extended quantum Hall family across various eigenspinor bases. Particularly, we reveal the emergence of a non-Abelian interface even when the bulks are Abelian, which allows the topologically trivial engineering of topologically protected edge states. We also define the braid group for pseudospin observables, demonstrating non-Abelian braiding operations and the Yang–Baxter relations. Our results pave the way for realizing a reconfigurable testbed for a wide class of Abelian and non-Abelian topological phenomena and braiding operations.

## Introduction

A versatile building block for reconfigurable unitary operations lies at the heart of classical[1,2] and quantum[3] physics. In programmable photonic circuits (PPCs)[2,4], which serve as pivotal frameworks for executing high-level functionalities with light, the standard building block is a reconfigurable universal SU(2) gate. Systematic assembly[1] of these gates enables higher-dimensional unitary operations essential for wave manipulations[5], matrix calculations[6], quantum computations[7], and Hamiltonian emulations[8-10]. A variety of proposals have been put forward to enhance the performance of unitary building blocks in terms of scalability[11,12], fidelity[13,14], speed[15], and energy efficiency[16].

Unitary operations play a crucial role also in the emerging field of non-Abelian physics[17-19]. Because symmetries and their associated gauge invariances are characterized by unitary operators, the noncommutative nature of internal symmetries in non-Abelian systems requires matrix-valued gauge fields within noncommutative unitary groups U($N$>1). In photonics, the successful implementation of non-Abelian gauge fields has been realized through both static[20-28] and reconfigurable platforms[29,30], utilizing anisotropic materials[20-22], metamolecules[23], coupled waveguides[24-28], nonreciprocal interferometers[29], and frequency-synthetic dimensions[30]. When evaluating PPCs as reconfigurable unitary operators[2,4] and considering recent progress in integrating PPCs into lattices with Abelian scalar gauge fields[8,9], the next challenge lies in devising a versatile building block for reconfigurable, universal, and lattice-compatible non-Abelian matrix-valued gauge fields.

Here we propose programmable photonic building blocks and their lattice assembly for exploring both Abelian and non-Abelian physics and their hybrids. The building block consists of two travelling-wave resonators coupled via a nonreciprocal loop coupler that provides a matrix-valued gauge field. Because the gauge field is adjusted by tuning phase shifters in the coupler, a periodic arrangement of these building blocks forms a reconfigurable spinor lattice, enabling the emulation of multiple topological phases, including both Abelian and non-Abelian phenomena. As a representative example, we implement an isospectral family of Abelian topological phenomena, such as the quantum Hall effect (QHE), and quantum spin Hall effects (QSHEs) for different bases of eigenspinors. Particularly, moving beyond conventional studies on non-Abelian topological bulks[17,18], we introduce the concept of a non-Abelian interface, which is characterized by noncommutative gauge-field distributions that emerge only at the interface between isospectral Abelian QSHE lattices. At the interface, we reveal the coexistence of topologically nontrivial edge states and their topologically trivial hybridizations, which leads to the reopening of bandgaps. We also visit another branch in non-Abelian physics, demonstrating the classical emulation of non-Abelian braiding[17-19,24,27,31] with our spinor lattices. Our theoretical results not only provide a foundational building block for non-Abelian and programmable topological photonics but also extend non-Abelian phenomena into the realm of interface physics.

## Results

### Programmable photonic spinor lattices

As a programmable platform for non-Abelian photonics, we employ a lattice composed of travelling-wave ring resonators[32] (Fig. 1a). Near the target operating frequency, the *m*th resonator supports degenerate pseudospin—counter-clockwise ($\psi_m^+$) and clockwise ($\psi_m^-$)—resonances (Fig. 1b), which constitute a pseudospinor state $\Psi_m = [\psi_m^+, \psi_m^-]^T$ in an expanded Hilbert space essential for matrix-valued gauge fields[17]. Although such lattices have been extensively studied in topological photonics[33-36], previous efforts have been limited to Abelian topological phenomena.

For example, to realize photonic QSHEs[33,34], the SU(2) link variable is implemented with a gauge field proportional to the Pauli $z$ matrix $\sigma_z$, which drives a $z$-axis rotation on the Bloch sphere of $\Psi_m$. Moreover, the influence of defects in resonators, which can mix pseudospins, has broadened the range of accessible SU(2) link variables, including $x$-axis rotations for modelling Rashba-like effects[33,37]. However, creating a reconfigurable lattice that encompasses both Abelian and non-Abelian physics remains a challenge.

In this context, we revisit the resonator lattice to develop a programmable building block for U(2) gauge fields. The core component is a waveguide loop coupler evanescently coupled to the resonators with a decay rate of $1/\tau$. The coupler consists of two nonreciprocal parts: an SU(2) gate, up to a global phase, and a global phase shifter (Fig. 1b). The first part mainly follows the design of a conventional SU(2) gate in programmable photonics[2,4], utilizing Mach-Zehnder interferometers and local phase shifters applied to the upper arm of the coupler. However, in our design, one of the local phase shifters implements a nonreciprocal phase shift (NRPS) $\pm 2\xi_L$, while the other phase shift $\eta_L$ remains reciprocal. The second part introduces the NRPSs, $\xi_G^F$ and $\xi_G^B$, globally in both arms. The lattice composed of the building blocks is governed by the following tight-binding Hamiltonian with matrix-valued gauge fields (Supplementary Note S1):

$$H = -\frac{1}{2\tau}\sum_{\langle m,n\rangle}\begin{pmatrix}\hat{a}_{\uparrow m}{}^{\dagger} & \hat{a}_{\downarrow m}{}^{\dagger}\end{pmatrix} e^{i\mu_{mn}} U_{mn}{}^{\dagger}\begin{pmatrix}\hat{a}_{\uparrow n}\\ \hat{a}_{\downarrow n}\end{pmatrix} + \text{H.c.} \tag{1}$$

where ($\hat{a}_{\uparrow m}{}^{\dagger}$, $\hat{a}_{\uparrow m}$) and ($\hat{a}_{\downarrow m}{}^{\dagger}$, $\hat{a}_{\downarrow m}$) are the pairs of the creation and annihilation operators for the counter-clockwise and clockwise pseudospin resonances, respectively, $\langle m,n\rangle$ denotes a pair of nearest-neighbour indices, $\mu_{mn}$ is the U(1) gauge field, and $U_{mn}$ is the SU(2) link variable, which is tailored solely by the local phase shifts $\xi_L$ and $\eta_L$, as follows:

$$U_{mn} = \begin{bmatrix} e^{-\frac{i}{2}(\eta_\mathrm{L}+\pi)} \sin\xi_\mathrm{L} & e^{+\frac{i}{2}(\eta_\mathrm{L}+\pi)} \cos\xi_\mathrm{L} \\ e^{-\frac{i}{2}(\eta_\mathrm{L}-\pi)} \cos\xi_\mathrm{L} & e^{+\frac{i}{2}(\eta_\mathrm{L}+\pi)} \sin\xi_\mathrm{L} \end{bmatrix}. \quad (2)$$

As shown in Eq. (2), the critical parameter for non-Abelian physics is the NRPS $\xi_\mathrm{L}$, which couples counter-clockwise and clockwise resonances, thereby allowing the implementation of non-Abelian U(2) gauge fields. We also note that the Abelian U(1) gauge field is determined by both local and global phase shifts, as $\mu_{mn} = \xi_\mathrm{G}^\mathrm{F} + \xi_\mathrm{L} + \eta_\mathrm{L}/2$, while the necessary $\xi_\mathrm{G}^\mathrm{B}$ is routinely determined by the other shifts, as $\xi_\mathrm{G}^\mathrm{B} = -\eta_\mathrm{L} - \xi_\mathrm{G}^\mathrm{F} - \pi$ (modulo $2\pi$). The utilization of a global phase for U(1) gauge fields shows the uniqueness of our coupler compared to SU(2) gates in programmable photonics[2,4], which are designed only up to a global phase.

The SU(2) gauge field $U_{mn}$ can be interpreted as a rotation on the spinor Bloch sphere during the coupling process: $U_{mn} = \exp(-i\alpha\mathbf{n}\cdot\boldsymbol{\sigma}/2)$, representing the $\alpha$-angle rotation about the **n**-axis, where $\boldsymbol{\sigma} = \sigma_x\mathbf{x} + \sigma_y\mathbf{y} + \sigma_z\mathbf{z}$ is the Pauli vector. Because $U_{mn}$ allows for complete rotations around the $y$- and $z$-axes with $2\pi$ rotation ranges (Methods and Supplementary Note S2), we adopt these two rotations as fundamental operation modes (Fig. 1c,d), which constitute a universal set for single-qubit gates[38]. For full-range operation modes, the coupler requires the phase shifts ranging from $0 \leq \xi_\mathrm{L} \leq \pi$ and $0 \leq \eta_\mathrm{L} \leq 2\pi$.

While the reciprocal part, $\eta_\mathrm{L}$, can be implemented using thermo-optical[39] or microelectromechanical[16] modulations, the realization of NRPS, $\xi_\mathrm{L}$, poses a challenge in terms of integration. Among various approaches for achieving NRPS[40-42], we employ a cerium-substituted yttrium iron garnet (Ce:YIG) silicon waveguide[40] to evaluate the real implementation of our design. Figures 1e and 1f show the cross-sectional views of the structure and eigenmode of the nonreciprocal waveguide operating at 1550 nm, respectively, which are analysed with the finite-difference-frequency-domain (FDFD) method in the commercial software Tidy3D[43]. Within the linear range of the relationship between the Faraday rotation and an external magnetic field in the Ce:YIG material[44,45], the structure provides a NRPS of 1.4 rad/mm for a Faraday rotation of 6000 deg/cm (Fig. 1g). Especially, due to the linearity in Fig. 1g, the employed waveguide allows for linear control of $\xi_L$ with the strength of the external magnetic field. We also design the other components of the building block using the finite-difference-time-domain (FDTD) method and FDFD method in the Tidy3D[43], including reciprocal phase shifters with thermo-optical modulation, Mach-Zehnder interferometers, and waveguide-resonator coupling (Supplementary Note S3). All components are designed to be compatible with the Ce:YIG-NRPS and are integrated within a 0.62 mm$^2$ footprint, experiencing a loss of 1.35 dB.

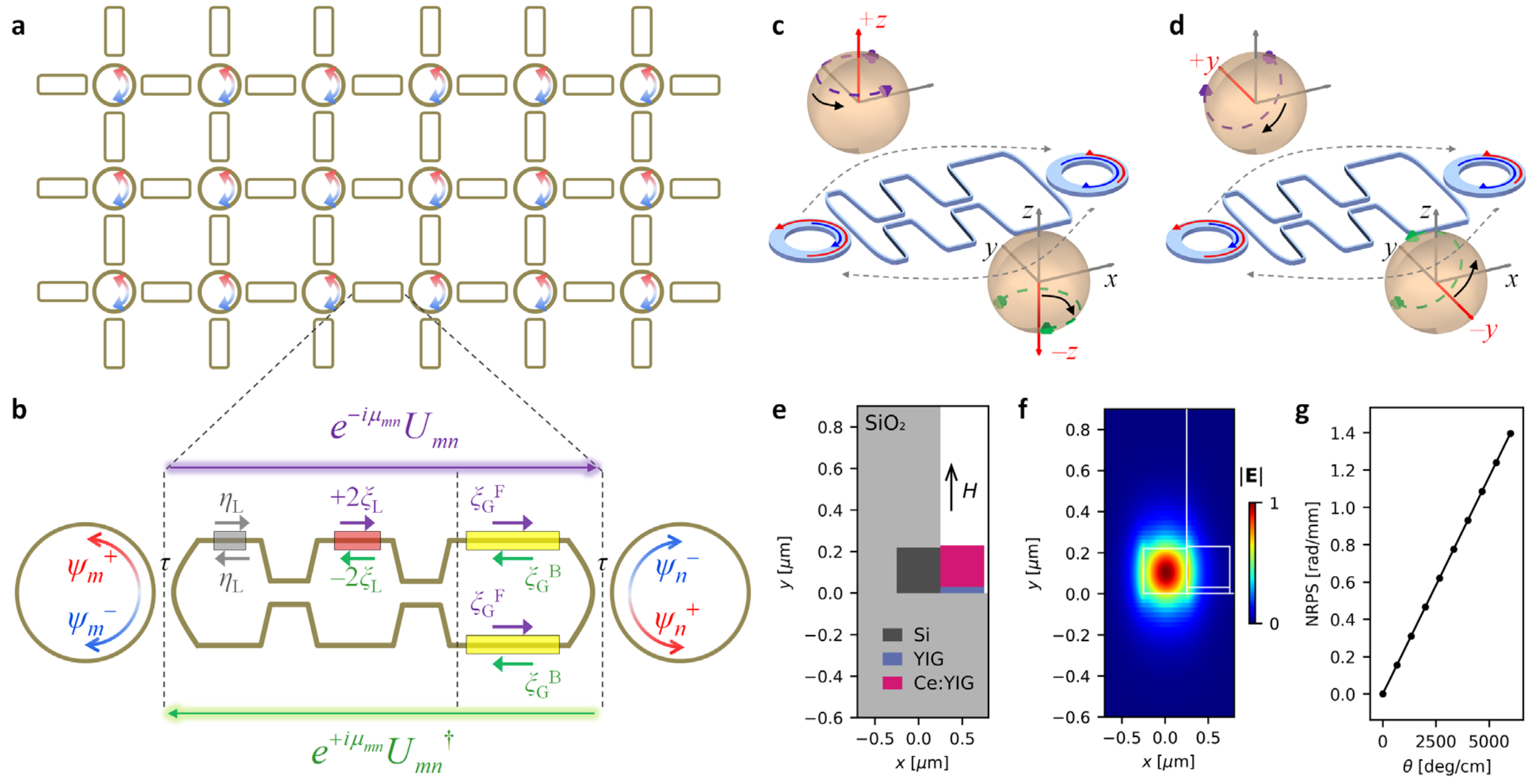

**Fig. 1. Programmable spinor lattices with U(2) gauge fields. a,** A square lattice composed of pseudospinor resonators. **b,** The building block of a spinor lattice. The grey, red, and yellow boxes represent local-reciprocal, local-nonreciprocal, and global-nonreciprocal phase shifters, respectively. Purple and green arrows indicate the forward and backward directions, respectively. **c,d,** Loop coupler operations for rotations around the $z$-axis (**c**) and $y$-axis (**d**). Grey dashed arrows illustrate the direction of coupling, accompanying the rotations depicted on the Bloch spheres. The Hermiticity of $H$ with $U_{mn} = U_{nm}^\dagger$ results in the same rotation angle but with opposite rotation axes

(red straight arrows) for opposite coupling directions. Red and blue circular arrows in **a-d** denote counter-clockwise and clockwise pseudospin resonances, respectively. **e,f,** Cross-sectional views of the nonreciprocal waveguide (**e**) and its eigenmode electric field distribution (**f**) at the 1550 nm wavelength. The design is motivated by the previous experimental study[40]. **g,** The NRPS of the waveguide as a function of the Faraday rotation in the Ce:YIG material. The result is calculated using the FDFD mode solver of Tidy3D[43].

**Isospectral Abelian topological lattices**

By assembling the designed building blocks, we develop a spinor lattice for the U(2)-flux generalization of the standard Harper-Hofstadter (HH) model[46,47] as the first example. The topological nature of the lattice is governed by the loop operator of each plaquette: $K = \mathcal{P}\Pi\exp(-i\mu)\exp(-i\alpha\mathbf{n}\cdot\boldsymbol{\sigma}/2)$, where $\mathcal{P}$ denotes the path-ordered product of matrix-valued link variables around a plaquette. The gauge field parameters $\mu$, $\alpha$, and $\mathbf{n}$ are defined individually for each coupler path in determining $K$. For the generalized HH model, the Hermiticity of the Hamiltonian leads to (see Methods)

$$K = \left[K_3^{(p,q+1)}\right]^\dagger \left[K_1^{(p+1,q+1)}\right]^\dagger K_3^{(p+1,q+1)} K_1^{(p+1,q)}, \tag{3}$$

where $K_\gamma^{(p,q)} \equiv \exp(-i\mu_\gamma^{(p,q)})U_\gamma^{(p,q)}$ ($\gamma = 1, 2, 3$, and $4$) denotes the $\gamma$-directional incident link variable to the $(p,q)$th resonator (Fig. 2a).

We emphasize that the loop operator $K$, which corresponds to the matrix-valued flux across each plaquette, fully encompasses the U(2) group due to the universality of fundamental operation modes[38]. Consequently, the spinor lattice allows for the programmable emulation of both universal Abelian and non-Abelian topological phenomena on a single platform, including non-Abelian topological lattices; for example, $K_1^{(p+1,q)} \sim \exp(-i\alpha_z\sigma_z/2)$ and $K_3^{(p,q)} \sim \exp(-i\alpha_y\sigma_y/2)$, which corresponds to the symmetric-gauge generalization of the HH model[30,48]. At this stage, we investigate the spinor lattices to generalize the HH model within an Abelian group, which is characterized by the scalar flux parameters $\mu$ and $\alpha/2$, and a single type of **n** around the entire lattice. Firstly, we develop the standard HH model on our spinor lattice, characterized by the U(1) gauge field. The gauge field is expressed as $K \equiv K_0 = \exp(-i\mu)\sigma_0$ in the loop operator representation, where $\sigma_0$ is the 2 × 2 identity matrix. Figures 2b and 2c show the traditional Hofstadter butterflies for this standard HH model, illustrating the gap Chern number $C$ and the spin gap Chern number $C_S$, respectively. Notably, the time-reversal symmetry of the system is broken in both local and global manners, as evidenced by NRPSs in each coupler and the same sign of $C_S$ for both pseudospins, which is the signature of the QHE.

In the same platform, we can program the lattice to achieve the isospectral partner of the HH model with the transformed basis of eigenspinors. For example, by tailoring the phase shifters, the standard QSHE with $K \equiv K_z = \exp(-i\alpha\sigma_z/2)$ can be achieved. As demonstrated in previous studies[47,49,50], the gap Chern number is zero in all regimes (Fig. 2d), while the spin gap Chern number is nonzero and has an opposite sign in each pseudospin resonance (Fig. 2e,f). We note that the unique form of the QSHE can also be achieved by setting $K \equiv K_y = \exp(-i\alpha\sigma_y/2)$. This Abelian SU(2) gauge field derives the coupling between pseudospin modes, imposing topological natures of the QSHE on the transformed spinor basis: $[1,+i]^T/2^{1/2}$ and $[1,-i]^T/2^{1/2}$. Additionally, both QSHEs achieve globally preserved time-reversal symmetry despite NRPSs, as evidenced by the zero gap Chern numbers. Therefore, the proposed platform allows for constituting an isospectral family of topological lattices, engineering eigenspinor bases and time-reversal symmetry. This engineering can be achieved dynamically in a given platform, by programming the distribution of coupler phase shifts. Up to now, we have demonstrated QHE and QSHE associated with Abelian gauge field physics as the vector **n** is the same throughout the medium. On the other hand, a spatially varying **n** will deviate from a purely Abelian picture and require a non-Abelian physics description.

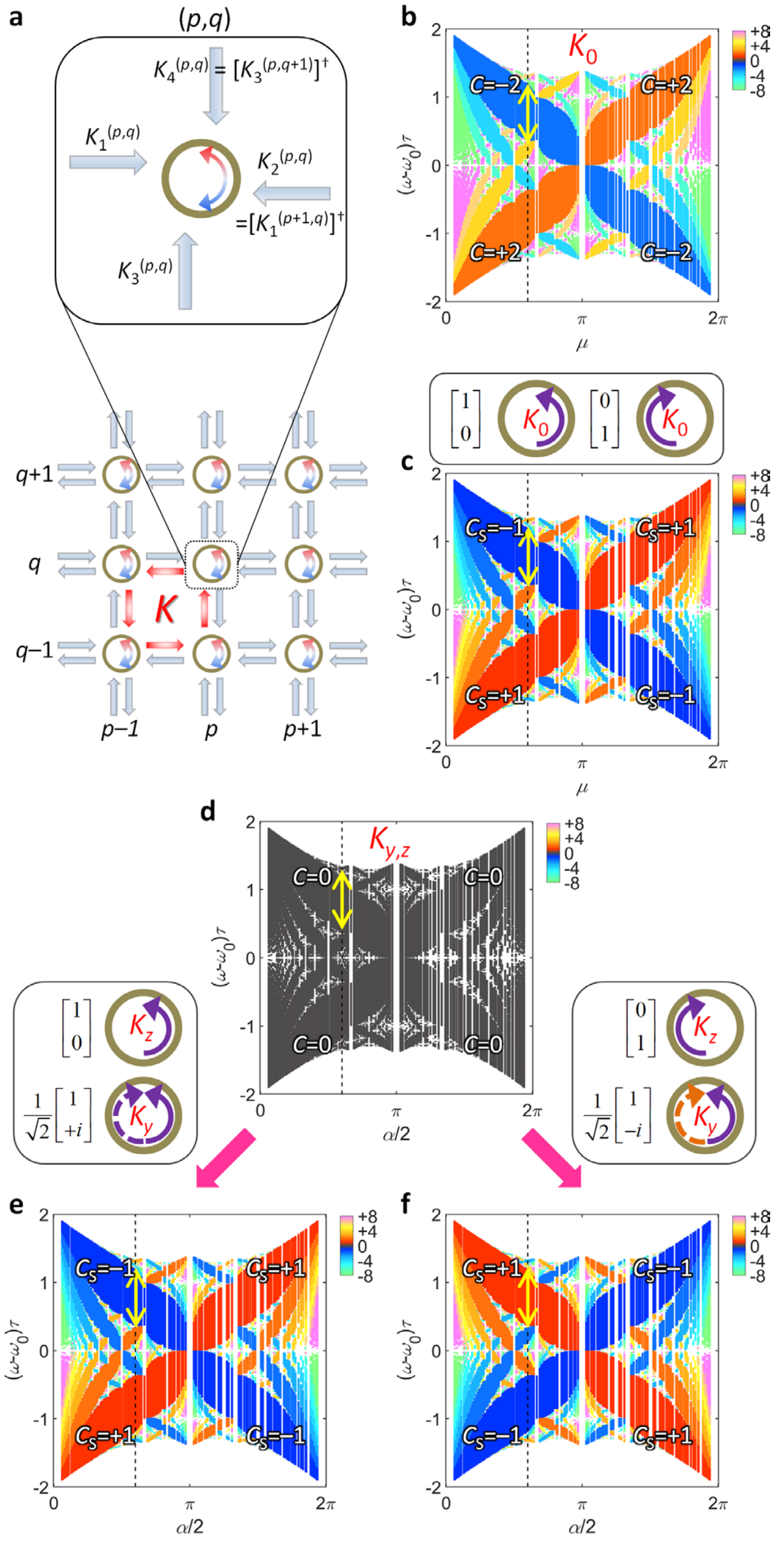

**Fig. 2. Isospectral Abelian topological lattices. a,** A schematic for the distribution of the link variables $K_\gamma^{(p,q)}$ around the ($p$,$q$)th unit cell, where $\gamma$ = 1, 2, 3, and 4 indicates the relative direction of the matrix-valued gauge field. The gauge fields are designed for spatial homogeneity and Hermiticity. **b,c,** QHE Hofstadter butterflies for gap Chern numbers $C$ (**b**) and spin gap Chern numbers $C_S$ (**c**) for the loop operator $K_0$. In QHE, $C_S$ are identical for both eigenspinors. **d-f,** QSHE

Hofstadter butterflies for $C$ (**d**) and $C_S$ (**e,f**) for the loop operators $K_y$ and $K_z$. $K_y$ and $K_z$ lead to different eigenspinor bases, while the cases of $[1,0]^T$ and $[1,+i]^T/2^{1/2}$ eigenspinors (or the cases of $[0,1]^T$ and $[1,-i]^T/2^{1/2}$ eigenspinors) lead to the same $C_S$. The calculation of butterflies is conducted using the conventional method of solving the eigenvalue problem within a $\Lambda_x \times \Lambda_y$ supercell in the first Brillouin zone[46], where $\Lambda_x$ and $\Lambda_y$ represent the periodicities along the $x$- and $y$-axes, respectively. Details of the butterfly visualization are described in Methods.

**Non-Abelian topological interfaces**

Utilizing the isospectral family of Abelian topological lattices, we explore unique non-Abelian phenomena observed at the interface between Abelian topological bulks, which we refer to as a non-Abelian interface. Figure 3a describes the interface between the upper and lower Abelian lattices, which have the loop operators $K_U$ and $K_L$, respectively. Following the discussion in Fig. 2, we focus on the loop operators $K_0$ (Fig. 3b), $K_y$ (Fig. 3c), and $K_z$ (Fig. 3d) for $K_{U,L}$. Considering the mirror symmetry across the interface, three types of interfaces can be implemented: ($K_U$, $K_L$) = ($K_0$, $K_y$), ($K_0$, $K_z$), and ($K_y$, $K_z$). In illustrating the eigenspinors and the corresponding band structures of these interfaces (Fig. 3e-g), we utilize the mixed colormaps for three spinor bases (Fig. 3h and Methods).

The first two types of interfaces, ($K_0$, $K_y$) and ($K_0$, $K_z$) (Fig. 3e,f), are classified as Abelian ones because $K_0 = \exp(-i\mu)\sigma_0$ commutes with any SU(2) link variables. By examining the Hofstadter butterflies with their spin gap Chern numbers (Fig. 2c,e,f), we show that the nontrivial topological interfaces in these Abelian configurations are achieved only for one of the spinors: $[1,-i]^T/2^{1/2}$ in Fig. 3e and $[0,1]^T$ in Fig. 3f. The following emergence of a topologically-protected edge state for a single spinor highlights the uniqueness of our Abelian topological interfaces ($K_0$, $K_y$) and ($K_0$, $K_z$), in stark contrast to representative edge states in topological photonics[47]. For example, in the QHE, which occurs at the interface between bulks with different U(1) gauge fields—denoted as the interface ($K_0$, $K_0'$) in our notation, where $K_0 \neq K_0'$—chiral edge states feature the same directional edge states for both pseudospins. Similarly, in the QSHE, which occurs at the interface between bulks characterized by different $z$-axis Bloch-sphere rotations—the interface ($K_z$, $K_z'$), where $K_z \neq K_z'$—helical edge states exhibit opposite directional edge states for pseudospins.

On the other hand, we explore the configuration ($K_y$, $K_z$), which introduces non-Abelian interface physics that offers a method for engineering topologically protected edge states (Fig. 3g). The configuration satisfies the non-Abelian condition only at the interface while maintaining Abelian bulks, because $K_y = \exp(-i\alpha\sigma_y/2)$ and $K_z = \exp(-i\alpha\sigma_z/2)$ leads to noncommutative loop products around the interface due to the nonzero commutation $[\sigma_y,\sigma_z] = 2i\sigma_x$. In this configuration, which cannot be characterized with scalar-valued fluxes, two nontrivial topological interfaces are achieved for different pairs of the coupled spinors from distinct bases: the pair of $[1,-i]^T/2^{1/2}$ and $[1,0]^T$, and the pair of $[1,+i]^T/2^{1/2}$ and $[0,1]^T$ (Fig. 3g). Because these interfaces provide an opposite spatial arrangement of spin gap Chern numbers (red boxes in Fig. 3g), the induced helical edge states have opposite signs of group velocity, which is consistent with the globally preserved time-reversal symmetry. Furthermore, different basis representations for eigenspinors in the $K_y$ and $K_z$ regions lead to coupling within the pairs of states that have the same spin gap Chern number (black solid arrows in Fig. 3g). This topologically trivial coupling results in the hybridization of helical edge states and the band anticrossing (red solid arrows in Fig. 3g), eventually restoring bandgaps, which is the unique phenomenon of non-Abelian topological interfaces. Another unique property of non-Abelian topological interfaces—the spatial distribution of edge-state spinors—is discussed in Supplementary Note S4.

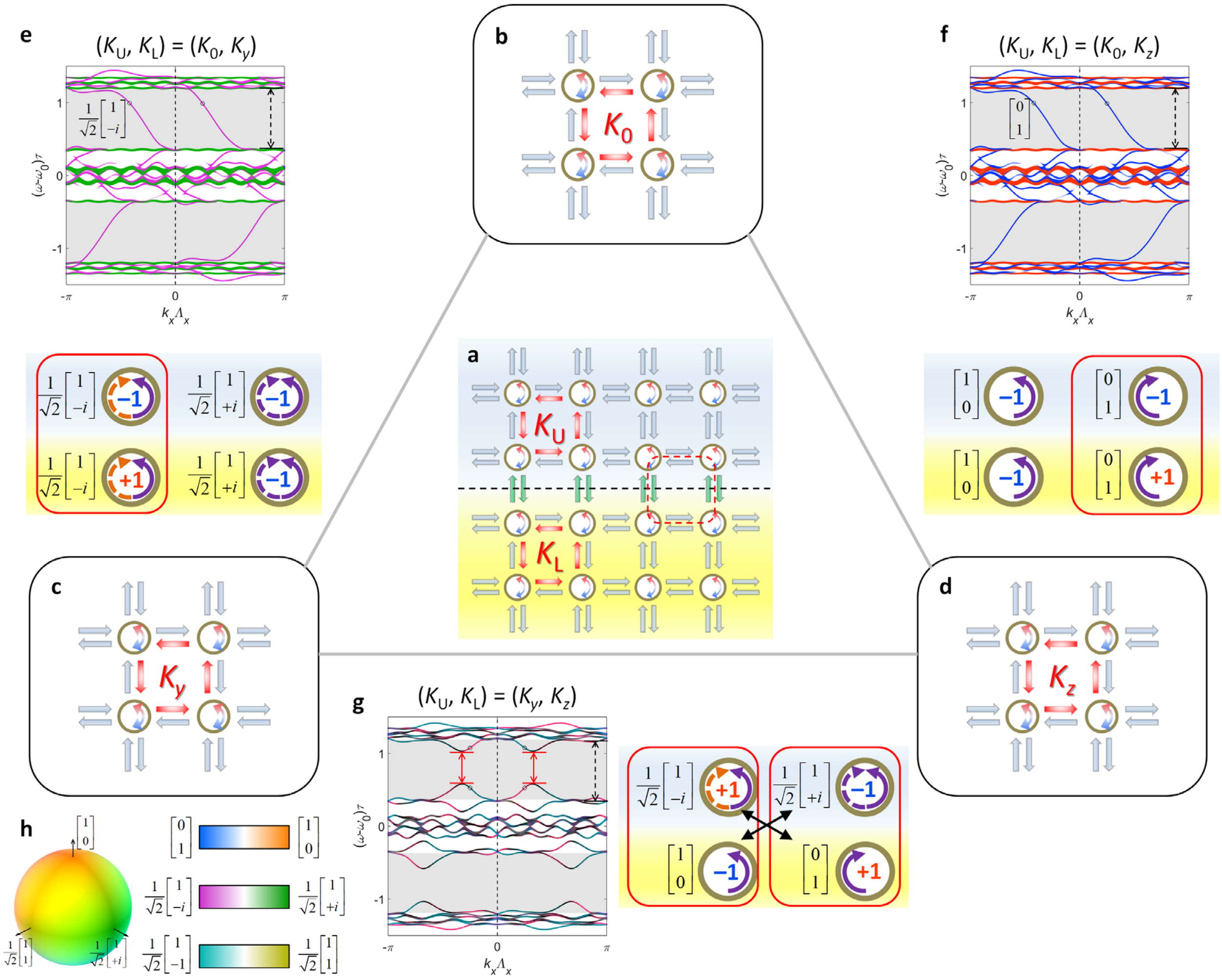

**Fig. 3. Abelian and non-Abelian topological interfaces. a,** A schematic for the interface (black dashed line) between the lattices with loop operators $K_U$ and $K_L$. Green arrows indicate the coupling with the zero gauge field. Red dashed box denotes the interface loop. **b-d,** Topological lattices with $K_0$ (**b**), $K_y$ (**c**), and $K_z$ (**d**) loop operators. **e-g,** Band diagrams projected onto the $k_x$-axis, where $k_x$ denotes the $x$-axis component of the reciprocal lattice vector, and the corresponding topologically nontrivial interface configurations for the Abelian cases, ($K_0$, $K_y$) (**e**) and ($K_0$, $K_z$) (**f**), and the non-Abelian case, ($K_y$, $K_z$) (**g**). The circles with coloured arrows represent the eigenspinors and the corresponding spin gap Chern numbers. **h,** Colormap mixing for illustrating spinor bases on the Bloch sphere (Methods), which is applied to the bands of (**e-g**). Black dashed arrows in (**e-g**) depict the original bandgap of the bulk lattices. Red solid arrows in (**g**) highlight the bandgap reopening. The fluxes applied to each bulk lattice correspond to the cases depicted with black dashed lines in the bands of Fig. 2b-f, which are characterized by the supercells of $\Lambda_x = 1$ and $\Lambda_y =$

10 with the rational-number fluxes $\mu = 0.6\pi$ and $\alpha/2 = 0.6\pi$. Details of calculating band structures using the supercell configuration are described in Methods.

**Non-Abelian resonant braiding**

Using our building block, we demonstrate the photonic analogy of non-Abelian braiding[17,18,24,27,31,51], which has attracted substantial attention in describing the exchange of anyons as the gate operation for quantum computation[19]. To model quasiparticles with the degeneracy essential for non-Abelian braiding[19,27], we assign spin observables—the Bloch vector components of pseudospin resonances—to the particles: $S_j^m = \langle\Psi_m|\sigma_j|\Psi_m\rangle$ ($j = x$, $y$, and $z$) in the $m$th resonator. The evolution of particle exchanges is realized by the coupling process (Fig. 4a), which accompanies rotation operations achieved with the loop coupler. As the generators of the braid group describing the particle exchanges, we employ fundamental operation modes of the building block: the rotation operations $U_y = \exp(-i\alpha_y\sigma_y/2)$ and $U_z = \exp(-i\alpha_z\sigma_z/2)$, and their inverses $U_y^{-1}$ and $U_z^{-1}$ (Fig. 4b). $U_{y,z}$ and $U_{y,z}^{-1}$ correspond to counterclockwise and clockwise exchanges, respectively.

By constructing a one-dimensional (1D) coupled-resonator lattice, the 2 + 1 space-time dimensions for non-Abelian anyons can be emulated by the two-dimensional (2D) surface of the Bloch sphere and the 1D resonant coupling along the lattice. While the direction of the resonant coupling determined by the incident and transmitted waves corresponds to the arrow of time (Fig. 4c), the particle trajectory over time is visualized as a strand[19], defining the strand $S_j$ as the series of $S_j^m$. Therefore, the lattice describes the temporal exchange of three particles via the evolution of $S_x$, $S_y$, and $S_z$, formulating the non-Abelian braid group $B_3$. Considering the generators of $B_3$, we sequence the strands as $S_y$, $S_x$, and $S_z$.

A rotation around the $y$- (or $z$-) axis induces a unitary interaction between the $z$- and $x$-spin observables (or the $x$- and $y$-spin observables), which is visualized by the braiding of the corresponding strands. Although the interaction can be freely adjusted by the rotation angles $\alpha_y$ and $\alpha_z$, the allowed operations for $B_3$ are restricted by the criteria of the braid group—namely, far commutativity and the Yang–Baxter relation[52]. While the far commutativity is automatically satisfied within $B_3$, the Yang-Baxter relation requires $U_yU_zU_y = U_zU_yU_z$. Moreover, non-Abelian braiding enforces additional condition $U_yU_z \neq U_zU_y$. When restricting our discussion to the rotation angles of $0 \leq \alpha_{y,z} < 2\pi$, the allowed rotation operations that satisfy the listed criteria are achieved solely with $(\alpha_y,\alpha_z) = (\pi/2,\pi/2)$ and $(3\pi/2,3\pi/2)$.

Figure 4c-f demonstrates the non-Abelian braiding group $B_3$ realized with our resonant building block at $(\alpha_y,\alpha_z) = (\pi/2,\pi/2)$. Because a unitary coupling between two resonators corresponds to a generator, the non-abelian and the Yang–Baxter relations are demonstrated using three (Fig. 4c,d) and four (Fig. 4e,f) coupled resonators, respectively. The initial and final states of the particles, or spin observables, are excited and measured through waveguide couplings to the resonators, respectively, which determines the arrow of time.

Figures 4d and 4f show the observation of the braid-group criteria in the transmission spectra. The results indicate that the criteria are satisfied throughout the entire spectra. For example, the non-Abelian nature accounts for the distinction between $U_zU_y$: $S_y = +1 \rightarrow S_x = -1$ and $U_yU_z$: $S_y = +1 \rightarrow S_z = +1$. Similarly, the identity of $U_yU_zU_y = U_zU_yU_z$ is confirmed by their perfect overlay in the transmitted spin observables. Despite the valid criteria in the entire spectra, complete conservation of the strands is achieved at the resonant-tunnelling frequencies, which correspond to perfect transmission. We also note that the braiding geometry of the strands is programmable through the control of phase shifts in the loop couplers between resonators.

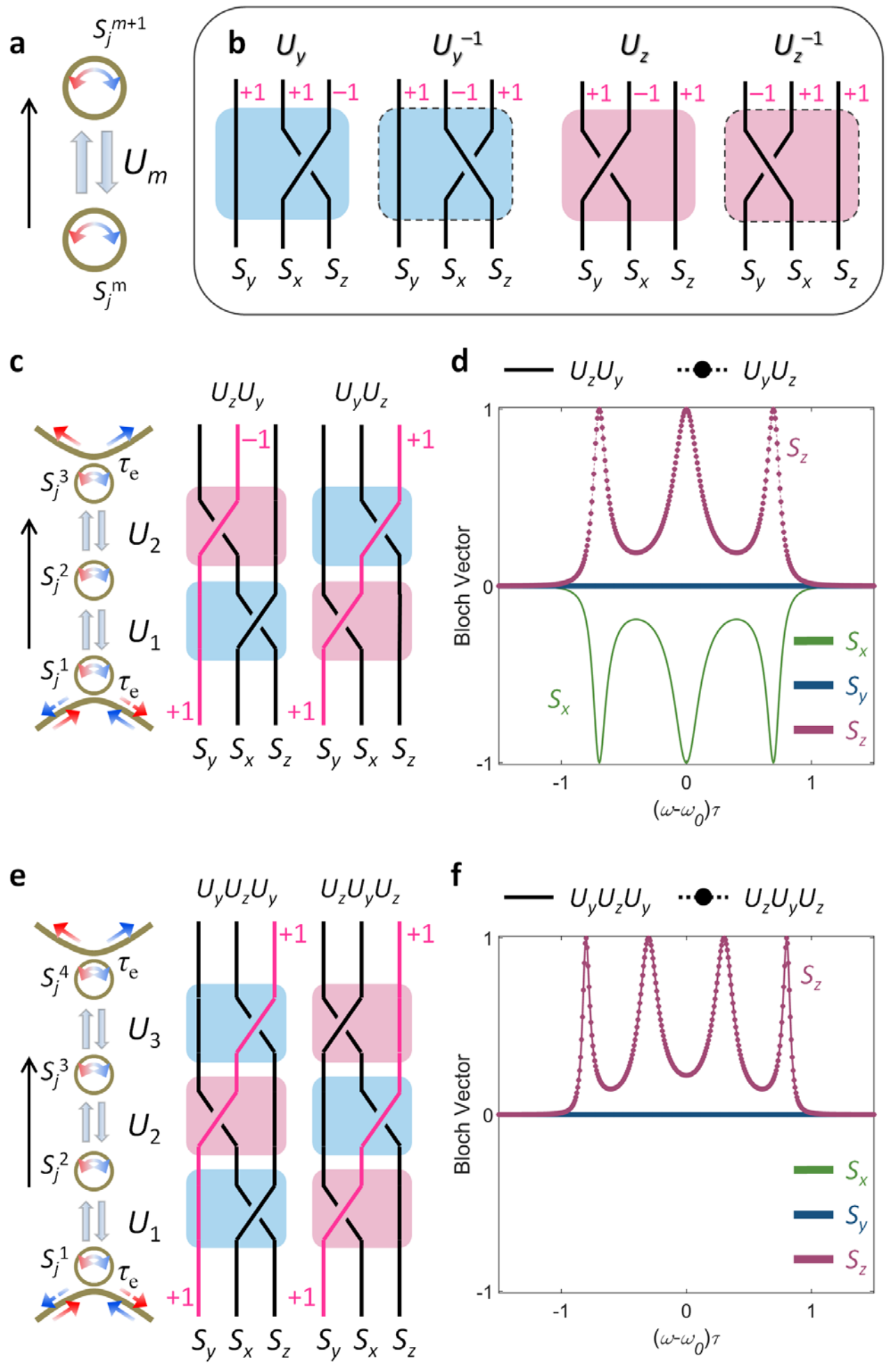

**Fig. 4. Programmable non-Abelian braiding. a,** The building block for emulating particle trajectories through resonator couplings. The pseudospin resonances, characterized by spin observables $S_j^m$ in the $m$th resonator, are coupled via a unitary link variable $U_m$ configured with the loop coupler. **b,** Generators for the braid group $B_3$. The number ±1 denotes the particle state after braiding when $S_j = 1$ is excited. **c-f,** The non-Abelian braid group criteria: **c,d,** non-Abelian condition, and **e,f,** Yang-Baxter relation. **c** and **e** show the 1D resonator lattices for demonstrating the criteria and the corresponding braid operations. The external waveguides are coupled to the resonators at the boundaries with a decay rate of $1/\tau_e$ to excite and measure the input and output spin observables, respectively. The cases of $(U_1,U_2) = (U_y,U_z)$ and $(U_z,U_y)$ are compared in **c** and **d**, and the cases of $(U_1,U_2,U_3) = (U_y,U_z,U_y)$ and $(U_z,U_y,U_z)$ are compared in **e** and **f**. **d** and **f** show the transmission spectra of the Bloch vector components through the lattices of **c** and **e**, respectively. $\tau_e = 4\tau$ in **c-f**. Black solid arrows in **a,c,e** denote the arrow of time. All the other parameters are the same as those in Figs. 2 and 3.

## Discussion

Due to the programmability of the proposed platform, abundant design freedom in time-reversal symmetry, bulk-edge configuration, and braiding operations can be utilized for reconfigurable light manipulation. For example, the demonstrated isospectral quantum Hall family allows for the control of time-reversal symmetry across the entire system, while each local component—nonreciprocal phase shifters—exhibits broken time-reversal symmetry. Abelian and non-Abelian interface states can be dynamically configured through the manipulation of a set of phase shifters within the subregion of the system. Reconfigurable braiding operations enable topologically nontrivial transitions between different knots and links. Such dynamical nature suggests the extension of the recently emerging field of time-varying photonics[53] into non-Abelian topological phenomena. We also note that time-varying modulations allow for magnetic-free realizations of our building blocks using synthetic magnetic fields[41] or acousto-optic modulation[42].

Substantial issues still remain from both fundamental and application perspectives. Reviewing studies on different dimensional defects in photonic crystals[54], extending research from the interface to point non-Abelian configurations could emerge as a future topic. The interfaces between non-Abelian topological lattices will offer substantial degrees of freedom due to their complex mothlike spectra[30,48,55]. Additionally, given the robustness of topological protection against external influences, the potential impact of non-Abelian interface physics could lie in enabling gate operations for topologically nontrivial states, as demonstrated in our gap reopening example. We also note that our braid group characterized by resonant, discretized, spin-observable, and spectral realizations enables novel forms of knots and links of photonic states, which will be in sharp contrast to previous approaches based on propagating, adiabatic, and spatial-mode realizations using waveguide arrays[24,27].

In conclusion, we have developed a building block for non-Abelian photonics. The reconfigurability of this building block allows for realizing isospectral quantum Hall family and constructing the interfaces between them. The non-Abelian interface exhibits unique characteristics that are sharply distinct from Abelian ones, demonstrating the coexistence of topologically trivial and nontrivial effects. The 1D lattice of our building blocks also allows for the non-Abelian braiding group for pseudospin observables. Our findings not only provide versatile platforms for exploring both Abelian and non-Abelian physics but also expand the scope of non-Abelian topological photonics into interface physics.

## Methods

**System parameters for rotations around the *y*- and *z*-axes.** By comparing Eq. (2) with $U_{mn}$ = $\exp(-i\alpha\mathbf{n}\cdot\boldsymbol{\sigma}/2)$, which represents the rotation by angle $\alpha$ about the **n**-axis, the required local phase shifts $\xi_L$ and $\eta_L$ can be determined from the given $\alpha$ and **n**, and vice versa. In this Methods section, we specifically address rotations about **n** = ±**y** and ±**z**, while more general cases are explored in Supplementary Note S2. It is well known that the Taylor expansion of $\exp(-i\alpha\mathbf{n}\cdot\boldsymbol{\sigma}/2)$ leads to the following expression[38]:

$$U_{mn} = e^{-i\frac{\alpha}{2}\mathbf{n}\cdot\boldsymbol{\sigma}} = \sigma_0 \cos\frac{\alpha}{2} - i\mathbf{n}\cdot\boldsymbol{\sigma}\sin\frac{\alpha}{2}, \tag{4}$$

where $\sigma_0$ is the 2 × 2 identity matrix. Equation (2) can also be expressed using the Pauli vector $\boldsymbol{\sigma}$:

$$U_{mn} = -\left(\sin\frac{\eta_L}{2}\sin\xi_L\right)\sigma_0 - i\left[\left(-\cos\frac{\eta_L}{2}\cos\xi_L\right)\sigma_x + \left(\sin\frac{\eta_L}{2}\cos\xi_L\right)\sigma_y + \left(\cos\frac{\eta_L}{2}\sin\xi_L\right)\sigma_z\right]. \tag{5}$$

For **n** = ±**z**, the coefficients of the Pauli $x$ and $y$ matrices should be zero, leading to

$$\xi_{\mathrm{L}} = \frac{\pi}{2}, \qquad \eta_{\mathrm{L}} = \pm\alpha - \pi + 4l_{\eta}\pi, \tag{6}$$

where $l_\eta$ is an integer number. Similarly, for $\mathbf{n} = \pm\mathbf{y}$, the coefficients of the Pauli $x$ and $z$ matrices should be zero, leading to

$$\xi_{\mathrm{L}} = \pm\frac{\alpha}{2} - \frac{\pi}{2} + 2l_{\xi}\pi, \qquad \eta_{\mathrm{L}} = \pi, \tag{7}$$

where $l_\xi$ is an integer number. We set $l_\eta = l_\xi = 0$, which determines the explicit ranges of local phase shifts $\xi_{\mathrm{L}}$ and $\eta_{\mathrm{L}}$ for the complete rotations about the $\pm y$- and $\pm z$-axes. The necessary phase shifts for each coupler operation mode are shown in Supplementary Note S2, requiring $0 \le \xi_{\mathrm{L}} \le \pi$ and $0 \le \eta_{\mathrm{L}} \le 2\pi$ to achieve a universal U(2) flux across a plaquette.

**Butterfly calculation.** In calculating the Hofstadter butterflies in Fig. 2b-f, the loop operator $K$ along the path $(p,q) \rightarrow (p+1,q) \rightarrow (p+1,q+1) \rightarrow (p,q+1)$ in Fig. 2a is defined by

$$K = K_4^{(p,q)} K_2^{(p,q+1)} K_3^{(p+1,q+1)} K_1^{(p+1,q)}. \tag{8}$$

To satisfy the Hermitian condition of the Hamiltonian, it is necessary that $K_2^{(p,q+1)} = [K_1^{(p+1,q+1)}]^\dagger$ and $K_4^{(p,q)} = [K_3^{(p,q+1)}]^\dagger$, which lead to Eq. (3). In Fig. 2, we consider three types of Abelian loop operators, $K_0 = \exp(-i\mu)$, $K_z = \exp(-i\alpha\sigma_z/2)$, and $K_y = \exp(-i\alpha\sigma_y/2)$, which can be collectively expressed as $K_d = \exp(-iv\sigma_d)$ ($v = \mu$ for $d = 0$, and $v = \alpha/2$ for $d = y$ and $z$). Similar to the Landau gauge[56], we set $K_1^{(p,q)} = \exp(iq\lambda_1\sigma_d)$ and $K_3^{(p,q)} = \exp(-ip\lambda_3\sigma_d)$, resulting in the homogeneous loop operator $K_d = \exp[-i(\lambda_1+\lambda_3)\sigma_d]$ around the entire lattice.

To satisfy the Bloch boundary conditions for the band calculation, $\lambda_1$ and $\lambda_3$ are set to be rational numbers. We calculate the band necessary for obtaining the butterflies when both $\lambda_1$ and $\lambda_3$ are irreducible fractions with denominators no greater than the natural number $B = 33$, and the sum of $\lambda_1$ and $\lambda_3$ lies between 0 and $2\pi$. The gap Chern number and spin gap Chern number are calculated according to their conventional definitions[49,57], while the spinor bases for these quantities are defined differently for $K_y$ and $K_z$, as $\{[1,1]^T, [0,1]^T\}$ for $K_z$, and $\{[1,+i]^T/2^{1/2}, [1,-i]^T/2^{1/2}\}$ for $K_y$.

**Band colouring.** Because the fundamental unit of our platform is a two-level system—travelling-wave resonator—comprising pseudospin resonances, the state of an eigenspinor at a specific resonator can be represented on the Bloch sphere. For a spinor state $|\Psi\rangle$, where $\langle m|\Psi\rangle = \Psi_m = [\psi_m^+, \psi_m^-]^T$ for the resonator lattice site operator $|m\rangle$, the state on the Bloch sphere is defined by three scalar degrees of freedom in Cartesian coordinates: $S_x = \langle\Psi|\sigma_x|\Psi\rangle$, $S_y = \langle\Psi|\sigma_y|\Psi\rangle$, and $S_z = \langle\Psi|\sigma_z|\Psi\rangle$. These scalars represent the $x$-, $y$-, and $z$-axis components of the position on the Bloch sphere, respectively, which characterize three pairs of eigenspinor bases. We apply corresponding colour maps to these bases as shown in Fig. 3h. The state of an eigenspinor at a specific band point ($k_x$, $k_y$, $\omega$), which will be located on the Bloch sphere, is visualized through additive colour mixing, as depicted in the coloured Bloch sphere of Fig. 3h.

**Supercell analysis for interface bands.** The results in Fig. 3 for interface physics are obtained from the lattice configurations using $K_0 = \mathcal{P}\Pi\exp(-i\mu_{mn}) = \exp(-i\mu)$, $K_y = \mathcal{P}\Pi\exp(-i\alpha_{mn}\sigma_y/2) = \exp(-i\alpha\sigma_y/2)$, and $K_z = \mathcal{P}\Pi\exp(-i\alpha_{mn}\sigma_z/2) = \exp(-i\alpha\sigma_z/2)$, where $\mu = \alpha/2 = 0.6\pi$. These flux values are implemented in the coupled-resonator supercells with $\Lambda_x = 1$ and $\Lambda_y = 10$. Consequently, for the calculation of the band structures shown in Fig. 3e-g, we apply the supercells of $\Lambda_x = 1$ and $\Lambda_y = 20$, placing the interface at the centre of the cell. We solve the eigenvalue problem across the first Brillouin zone, defined by $-\pi \leq k_x\Lambda_x \leq +\pi$ and $-\pi \leq k_y\Lambda_y \leq +\pi$, where $k_x$ and $k_y$ are the $x$- and $y$-components of the reciprocal lattice vector, respectively. To focus on the interface modes, we filter out the modes near the upper and lower boundaries by applying a proper weighting to the optical intensity of eigenmodes. The remaining bands overlap for different values of $k_y$. It is worth noting that the edge states are almost invariant with respect to $k_y$.

## Data availability

The data that support the plots and other findings of this study are available from the corresponding author upon request.

## Code availability

All code developed in this work will be made available upon request.

## Acknowledgements

We acknowledge financial support from the National Research Foundation of Korea (NRF) through the Basic Research Laboratory (No. RS-2024-00397664), Innovation Research Center (No. RS-2024-00413957), Young Researcher Program (No. 2021R1C1C1005031), and Midcareer Researcher Program (No. RS-2023-00274348), all funded by the Korean government (MSIT). This work was supported by Creative-Pioneering Researchers Program and the BK21 FOUR program of the Education and Research Program for Future ICT Pioneers in 2024, through Seoul National University. This work was also supported by the 2024 Research Fund of the University of Seoul for Xianji Piao. We also acknowledge an administrative support from SOFT foundry institute.

## Author contributions

S.Y., X.P., and N.P. conceived the project idea, which bridges programmable photonics with non-Abelian topological phenomena. X.P. and G.K. designed a building block for non-Abelian gauge fields using both analytical and numerical methods. S.Y., G.K., and J.L. analysed the topological phenomena and braiding groups, while S.Y. and N.P. interpreted the results in terms of non-Abelian physics. All authors contributed to the discussion of the results and the writing of the manuscript.

## Competing interests

A provisional patent application (KR Prov. App. 10-2024-0119460) has been filed by Seoul National University. The inventors include G.K., X.P., N.P., and S.Y. The application, which is pending, contains proposals of programmable photonic circuits for non-Abelian gauge fields.

## Additional information

**Correspondence and requests for materials** should be addressed to X.P., N.P., or S.Y.

## Figure Legends

**Fig. 1. Programmable spinor lattices with U(2) gauge fields. a,** A square lattice composed of pseudospinor resonators. **b,** The building block of a spinor lattice. The grey, red, and yellow boxes represent local-reciprocal, local-nonreciprocal, and global-nonreciprocal phase shifters, respectively. Purple and green arrows indicate the forward and backward directions, respectively. **c,d,** Loop coupler operations for rotations around the *z*-axis (**c**) and *y*-axis (**d**). Grey dashed arrows illustrate the direction of coupling, accompanying the rotations depicted on the Bloch spheres. The Hermiticity of $H$ with $U_{mn} = U_{nm}^\dagger$ results in the same rotation angle but with opposite rotation axes (red straight arrows) for opposite coupling directions. Red and blue circular arrows in **a-d** denote counter-clockwise and clockwise pseudospin resonances, respectively. **e,f,** Cross-sectional views of the nonreciprocal waveguide (**e**) and its eigenmode electric field distribution (**f**) at the 1550 nm wavelength. The design is motivated by the previous experimental study[40]. **g,** The NRPS of the waveguide as a function of the Faraday rotation in the Ce:YIG material. The result is calculated using the FDFD mode solver of Tidy3D[43].

**Fig. 2. Isospectral Abelian topological lattices. a,** A schematic for the distribution of the link variables $K_\gamma^{(p,q)}$ around the $(p,q)$th unit cell, where $\gamma = 1, 2, 3$, and 4 indicates the relative direction of the matrix-valued gauge field. The gauge fields are designed for spatial homogeneity and Hermiticity. **b,c,** QHE Hofstadter butterflies for gap Chern numbers $C$ (**b**) and spin gap Chern numbers $C_S$ (**c**) for the loop operator $K_0$. In QHE, $C_S$ are identical for both eigenspinors. **d-f,** QSHE Hofstadter butterflies for $C$ (**d**) and $C_S$ (**e,f**) for the loop operators $K_y$ and $K_z$. $K_y$ and $K_z$ lead to different eigenspinor bases, while the cases of $[1,0]^T$ and $[1,+i]^T/2^{1/2}$ eigenspinors (or the cases of $[0,1]^T$ and $[1,-i]^T/2^{1/2}$ eigenspinors) lead to the same $C_S$. The calculation of butterflies is conducted using the conventional method of solving the eigenvalue problem within a $\Lambda_x \times \Lambda_y$ supercell in the first Brillouin zone[46], where $\Lambda_x$ and $\Lambda_y$ represent the periodicities along the *x*- and *y*-axes, respectively. Details of the butterfly visualization are described in Methods.

**Fig. 3. Abelian and non-Abelian topological interfaces. a,** A schematic for the interface (black dashed line) between the lattices with loop operators $K_U$ and $K_L$. Green arrows indicate the coupling with the zero gauge field. Red dashed box denotes the interface loop. **b-d,** Topological lattices with $K_0$ (**b**), $K_y$ (**c**), and $K_z$ (**d**) loop operators. **e-g,** Band diagrams projected onto the $k_x$-

axis, where $k_x$ denotes the $x$-axis component of the reciprocal lattice vector, and the corresponding topologically nontrivial interface configurations for the Abelian cases, ($K_0$, $K_y$) (**e**) and ($K_0$, $K_z$) (**f**), and the non-Abelian case, ($K_y$, $K_z$) (**g**). The circles with coloured arrows represent the eigenspinors and the corresponding spin gap Chern numbers. **h,** Colormap mixing for illustrating spinor bases on the Bloch sphere (Methods), which is applied to the bands of (**e-g**). Black dashed arrows in (**e-g**) depict the original bandgap of the bulk lattices. Red solid arrows in (**g**) highlight the bandgap reopening. The fluxes applied to each bulk lattice correspond to the cases depicted with black dashed lines in the bands of Fig. 2b-f, which are characterized by the supercells of $\Lambda_x = 1$ and $\Lambda_y = 10$ with the rational-number fluxes $\mu = 0.6\pi$ and $\alpha/2 = 0.6\pi$. Details of calculating band structures using the supercell configuration are described in Methods.

**Fig. 4. Programmable non-Abelian braiding. a,** The building block for emulating particle trajectories through resonator couplings. The pseudospin resonances, characterized by spin observables $S_j^m$ in the $m$th resonator, are coupled via a unitary link variable $U_m$ configured with the loop coupler. **b,** Generators for the braid group $B_3$. The number ±1 denotes the particle state after braiding when $S_j = 1$ is excited. **c-f,** The non-Abelian braid group criteria: **c,d,** non-Abelian condition, and **e,f,** Yang-Baxter relation. **c** and **e** show the 1D resonator lattices for demonstrating the criteria and the corresponding braid operations. The external waveguides are coupled to the resonators at the boundaries with a decay rate of $1/\tau_e$ to excite and measure the input and output spin observables, respectively. The cases of $(U_1,U_2) = (U_y,U_z)$ and $(U_z,U_y)$ are compared in **c** and **d**, and the cases of $(U_1,U_2,U_3) = (U_y,U_z,U_y)$ and $(U_z,U_y,U_z)$ are compared in **e** and **f**. **d** and **f** show the transmission spectra of the Bloch vector components through the lattices of **c** and **e**, respectively. $\tau_e = 4\tau$ in **c-f**. Black solid arrows in **a,c,e** denote the arrow of time. All the other parameters are the same as those in Figs. 2 and 3.

# Supplementary Information for "Programmable lattices for non-Abelian topological photonics and braiding"

Gyunghun Kim[1], Jensen Li[2], Xianji Piao[3§], Namkyoo Park[4†], and Sunkyu Yu[1*]

[1]Intelligent Wave Systems Laboratory, Department of Electrical and Computer Engineering, Seoul National University, Seoul 08826, Korea

[2]Department of Engineering, University of Exeter, EX4 4QF, United Kingdom

[3]Wave Engineering Laboratory, School of Electrical and Computer Engineering, University of Seoul, Seoul 02504, Korea

[4]Photonic Systems Laboratory, Department of Electrical and Computer Engineering, Seoul National University, Seoul 08826, Korea

E-mail address for correspondence: [§]piao@uos.ac.kr, [†]nkpark@snu.ac.kr, [*]sunkyu.yu@snu.ac.kr

**Note S1. Programmable U(2) gauge fields in coupled resonator lattices**

Figure S1 illustrates a building block for a coupled resonator lattice that possesses programmable U(2) gauge fields. The $m$th resonator supports counter-clockwise ($\psi_m^+$) and clockwise ($\psi_m^-$) resonance modes, which comprise pseudospin modes for a spinor state $\Psi_m = [\psi_m^+, \psi_m^-]^T$. The $m$th and $n$th resonators are indirectly coupled through a waveguide loop, which is evanescently coupled to the resonators with a decay rate $1/\tau$. The fields inside the loop coupler, which are coupled with the mode of the $m$th resonator, are denoted by $v_{mI}^+$, $v_{mO}^+$, $v_{mI}^-$, and $v_{mO}^-$, where '+' and '–' correspond to the pseudospins of the coupled resonance modes, and 'I' and 'O' denote the input and output coupling with the resonator, respectively. We also apply the spinor representation to the loop coupler fields, represented as $V_{mI} = [v_{mI}^+, v_{mI}^-]^T$ and $V_{mO} = [v_{mO}^+, v_{mO}^-]^T$. The couplings between resonant and coupler spinor states are described by the temporal coupled mode equation[1]:

$$\begin{aligned} \frac{d\Psi_m}{dt} &= i\omega_m \Psi_m - \frac{1}{2\tau}\Psi_m + \sqrt{\frac{1}{\tau}} V_{mI}, \\ \frac{d\Psi_n}{dt} &= i\omega_n \Psi_n - \frac{1}{2\tau}\Psi_n + \sqrt{\frac{1}{\tau}} V_{nI}, \\ V_{mO} &= V_{mI} - \sqrt{\frac{1}{\tau}}\Psi_m, \\ V_{nO} &= V_{nI} - \sqrt{\frac{1}{\tau}}\Psi_n. \end{aligned} \tag{S1}$$

We note that waveguide loops have been extensively utilized in realizing U(1) Abelian gauge fields[2-5]. The key aspect of implementing U(2) gauge fields, including non-Abelian ones, using a waveguide loop lies in employing an SU(2) gate in programmable photonics[6-8], with nonreciprocal phase shifts (NRPSs). In detail, the waveguide loop in Fig. S1 consists of Regions I and II. Region I includes an SU(2) gate up to a global phase with two local phase shifters in the upper arm, $\eta_L$ and $\pm 2\xi_L$, while the second phase shifter is nonreciprocal for the forward ($+2\xi_L$) and backward ($-2\xi_L$) propagations. Region II provides a global NRPS to both arms with $\xi_G^F$ and $\xi_G^B$ for the forward and backward propagations, respectively. When a single trip through Regions I and II without phase shifter operations leads to the global phase evolutions $\gamma_{\mathrm{I}}$ and $\gamma_{\mathrm{II}}$, respectively, the transfer matrices through each region are shown as follows:

$$\begin{bmatrix} v_{\mathrm{U}}^{+} \\ v_{\mathrm{L}}^{-} \end{bmatrix} = T_{\mathrm{I}}^{+}\Big|_{\mathrm{UL}} \begin{bmatrix} v_{m\mathrm{O}}^{+} \\ v_{m\mathrm{O}}^{-} \end{bmatrix}, \text{ where } T_{\mathrm{I}}^{+}\Big|_{\mathrm{UL}} = e^{-i(\gamma_{\mathrm{I}}+\xi_{\mathrm{L}}+\eta_{\mathrm{L}}/2)} \begin{bmatrix} e^{-\frac{i}{2}(\eta_{\mathrm{L}}+\pi)}\sin\xi_{\mathrm{L}} & e^{+\frac{i}{2}(\eta_{\mathrm{L}}+\pi)}\cos\xi_{\mathrm{L}} \\ e^{-\frac{i}{2}(\eta_{\mathrm{L}}-\pi)}\cos\xi_{\mathrm{L}} & e^{+\frac{i}{2}(\eta_{\mathrm{L}}+\pi)}\sin\xi_{\mathrm{L}} \end{bmatrix},$$
$$\begin{bmatrix} v_{m\mathrm{I}}^{-} \\ v_{m\mathrm{I}}^{+} \end{bmatrix} = T_{\mathrm{I}}^{-}\Big|_{\mathrm{UL}} \begin{bmatrix} v_{\mathrm{U}}^{-} \\ v_{\mathrm{L}}^{+} \end{bmatrix}, \text{ where } T_{\mathrm{I}}^{-}\Big|_{\mathrm{UL}} = e^{-i(\gamma_{\mathrm{I}}-\xi_{\mathrm{L}}+\eta_{\mathrm{L}}/2)} \begin{bmatrix} e^{-\frac{i}{2}(\eta_{\mathrm{L}}-\pi)}\sin\xi_{\mathrm{L}} & e^{-\frac{i}{2}(\eta_{\mathrm{L}}-\pi)}\cos\xi_{\mathrm{L}} \\ e^{+\frac{i}{2}(\eta_{\mathrm{L}}+\pi)}\cos\xi_{\mathrm{L}} & e^{+\frac{i}{2}(\eta_{\mathrm{L}}-\pi)}\sin\xi_{\mathrm{L}} \end{bmatrix}, \tag{S2}$$
$$\begin{bmatrix} v_{n\mathrm{I}}^{+} \\ v_{n\mathrm{I}}^{-} \end{bmatrix} = T_{\mathrm{II}}^{+}\Big|_{\mathrm{UL}} \begin{bmatrix} v_{\mathrm{U}}^{+} \\ v_{\mathrm{L}}^{-} \end{bmatrix}, \text{ where } T_{\mathrm{II}}^{+}\Big|_{\mathrm{UL}} = e^{-i\left(\gamma_{\mathrm{II}}+\xi_{\mathrm{G}}^{\mathrm{F}}\right)}\sigma_0,$$
$$\begin{bmatrix} v_{\mathrm{U}}^{-} \\ v_{\mathrm{L}}^{+} \end{bmatrix} = T_{\mathrm{II}}^{-}\Big|_{\mathrm{UL}} \begin{bmatrix} v_{n\mathrm{O}}^{-} \\ v_{n\mathrm{O}}^{+} \end{bmatrix}, \text{ where } T_{\mathrm{II}}^{-}\Big|_{\mathrm{UL}} = e^{-i\left(\gamma_{\mathrm{II}}+\xi_{\mathrm{G}}^{\mathrm{B}}\right)}\sigma_0,$$

where $v_{\mathrm{U}}^{\pm}$ and $v_{\mathrm{L}}^{\pm}$ denote the loop coupler fields for the upper and lower arms, respectively, at the interface between Regions I and II, $T_{\mathrm{X}}^{\pm}|_{\mathrm{AB}}$ denotes the transfer matrix through Region 'X' (X = 'I' or 'II') for the ± direction, corresponding to the 'AB' order of the loop coupler arm in the column vectors (AB = 'UL' or 'LU' for upper (U) and lower (L) arms), and $\sigma_0$ is the 2 × 2 identity matrix (or the zeroth Pauli matrix). We note that the change in the 'AB' order, such as $[v_{\mathrm{U}}^{+},v_{\mathrm{L}}^{-}]^{\mathrm{T}} = T_{\mathrm{I}}^{+}|_{\mathrm{UL}}[v_{m\mathrm{O}}^{+},v_{m\mathrm{O}}^{-}]^{\mathrm{T}}$ and $[v_{\mathrm{L}}^{-},v_{\mathrm{U}}^{+}]^{\mathrm{T}} = T_{\mathrm{I}}^{+}|_{\mathrm{LU}}[v_{m\mathrm{O}}^{-},v_{m\mathrm{O}}^{+}]^{\mathrm{T}}$, leads to the basis change relation, $T_{\mathrm{X}}^{\pm}|_{\mathrm{AB}} = \sigma_x(T_{\mathrm{X}}^{\pm}|_{\mathrm{BA}})\sigma_x$, where $\sigma_x$ represents the Pauli $x$ matrix. Using this relation and defining $\gamma \triangleq \gamma_{\mathrm{I}} + \gamma_{\mathrm{II}}$, the transfer matrices between input and output coupler spinors are obtained as follows:

$$\begin{aligned} \mathrm{V}_{m\mathrm{I}} &= -e^{-i(\gamma+\eta_{\mathrm{L}}/2)}e^{-i\left(\xi_{\mathrm{G}}^{\mathrm{B}}-\xi_{\mathrm{L}}\right)}U_{mn}^{\dagger}\mathrm{V}_{n\mathrm{O}}, \\ \mathrm{V}_{n\mathrm{I}} &= +e^{-i(\gamma+\eta_{\mathrm{L}}/2)}e^{-i\left(\xi_{\mathrm{G}}^{\mathrm{F}}+\xi_{\mathrm{L}}\right)}U_{mn}\mathrm{V}_{m\mathrm{O}}, \end{aligned} \tag{S3}$$

where $U_{mn}$ is the SU(2) matrix presented in Eq. (2) in the main text.

In Eqs. (S1) and (S3), we substitute the coupler spinors $\mathrm{V}_{m\mathrm{I}}$, $\mathrm{V}_{m\mathrm{O}}$, $\mathrm{V}_{n\mathrm{I}}$, and $\mathrm{V}_{n\mathrm{O}}$ with the resonant spinors $\Psi_m$ and $\Psi_n$, which leads to the following coupled mode equation:

$$\frac{d}{dt}\begin{bmatrix}\Psi_m\\ \Psi_n\end{bmatrix}=\begin{bmatrix} i\omega_m\sigma_0-\frac{1}{2\tau}+\frac{1}{\tau}\frac{1}{1+e^{i\rho}} & +\frac{1}{\tau}\frac{1}{1+e^{i\rho}}e^{+i(\gamma+\eta_L/2)}e^{+i\left(\xi_G{}^F+\xi_L\right)}U_{mn}{}^{\dagger} \\ -\frac{1}{\tau}\frac{1}{1+e^{i\rho}}e^{+i(\gamma+\eta_L/2)}e^{+i\left(\xi_G{}^B-\xi_L\right)}U_{mn} & i\omega_n\sigma_0-\frac{1}{2\tau}+\frac{1}{\tau}\frac{1}{1+e^{i\rho}}\end{bmatrix}\begin{bmatrix}\Psi_m\\ \Psi_n\end{bmatrix}, \quad \text{(S4)}$$

where $\rho$ is the averaged phase evolution for the counter-clockwise and clockwise roundtrips along the loop coupler, as $\rho = 2\gamma + \eta_L + \xi_G{}^F + \xi_G{}^B$. By assigning the condition $\rho = 2b\pi$, where $b$ is an integer, Eq. (S4) is simplified as:

$$\frac{d}{dt}\begin{bmatrix}\Psi_m\\ \Psi_n\end{bmatrix}=\begin{bmatrix} i\omega_m\sigma_0 & +\frac{1}{2\tau}e^{+i(\gamma+\eta_L/2)}e^{+i\left(\xi_G{}^F+\xi_L\right)}U_{mn}{}^{\dagger} \\ -\frac{1}{2\tau}e^{+i(\gamma+\eta_L/2)}e^{+i\left(\xi_G{}^B-\xi_L\right)}U_{mn} & i\omega_n\sigma_0\end{bmatrix}\begin{bmatrix}\Psi_m\\ \Psi_n\end{bmatrix}. \quad \text{(S5)}$$

We also introduce the nonresonant condition of the loop coupler except for the phase shifts[2,5] as $2\gamma = (2b_R + 1)\pi$, where $b_R$ is an integer. While we select an even number $b_R$, we determine the nonreciprocal global phase shifts $\xi_G{}^F$ and $\xi_G{}^B$ with the target U(1) Abelian gauge field $\mu$ and the local phase shifts $\xi_L$ and $\eta_L$, as $\xi_G{}^F = 2b_F\pi - \xi_L - \eta_L/2 + \mu$ and $\xi_G{}^B = (2b_B - 1)\pi + \xi_L - \eta_L/2 - \mu$, where $b_F$ and $b_B$ are integers that are freely tunable according to the hardware design of the global phase shifter. Equation (S5) then becomes the following Hamiltonian form:

$$i\frac{d}{dt}\begin{bmatrix}\Psi_m\\ \Psi_n\end{bmatrix}=-\begin{bmatrix} \omega_m\sigma_0 & \frac{1}{2\tau}e^{+i\mu}U_{mn}{}^{\dagger} \\ \frac{1}{2\tau}e^{-i\mu}U_{mn} & \omega_n\sigma_0\end{bmatrix}\begin{bmatrix}\Psi_m\\ \Psi_n\end{bmatrix}. \quad \text{(S6)}$$

Equation (S6) demonstrates that the suggested building block enables the resonator coupling possessing the U(1) Abelian and SU(2) matrix-valued gauge fields, which allows for U(2) non-Abelian gauge fields. While the SU(2) gauge field is determined by local phase shifters $\xi_L$ and $\eta_L$, the U(1) gauge field is controlled with the global phase shifts $\xi_G{}^F$ and $\xi_G{}^B$. Equation (S6) derives the tight-binding Hamiltonian $H$ in Eq. (1) in the main text.

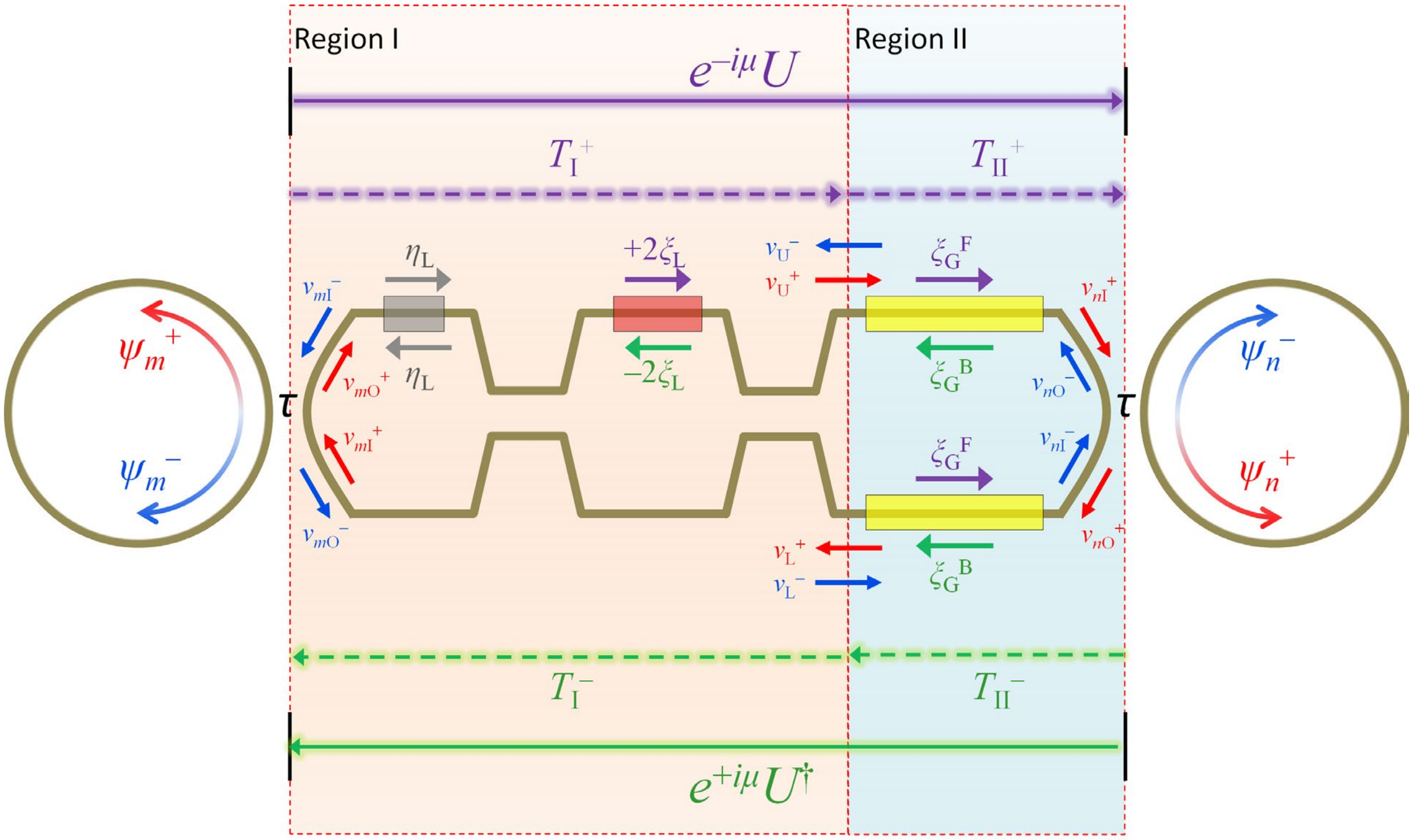

**Fig. S1. Spinor coupler with U(2) gauge fields.** The coupler links the *m*th and *n*th travelling-wave resonators. The grey coloured box represents the reciprocal phase shifter. Red and blue arrows denote the counter-clockwise and clockwise pseudospin resonances, respectively, as well as their interacting loop-coupler fields. The red and yellow boxes represent local and global nonreciprocal phase shifters, respectively. Purple and green arrows indicate the forward and backward directions, respectively.

**Note S2. Accessible rotation axes**

The relationship between the rotation operation, defined by **n** and $\alpha$, and the system parameters $\xi_L$ and $\eta_L$, is characterized by Eqs. (4) and (5) in the main text. As demonstrated in the main text and the Methods section, the proposed building block provides arbitrary rotations about **n** = ±**z** and **n** = ±**y** axes by following Eqs. (6) and (7) in the main text. Figures S2a and S2b show the necessary phase shifts for each coupler operation mode. When the rotation axes are fixed at **n** = +**z** and **n** = +**y**, complete rotations ($\pi \leq \alpha \leq 3\pi$) along the $z$- and $y$-axes can be achieved with the phase shifts ranging from $0 \leq \xi_L \leq \pi$ and $0 \leq \eta_L \leq 2\pi$. These degrees of freedom constitute a universal set of rotation gates for a spinor along the cell, thereby enabling a universal unitary loop operator $K$ in the main text.

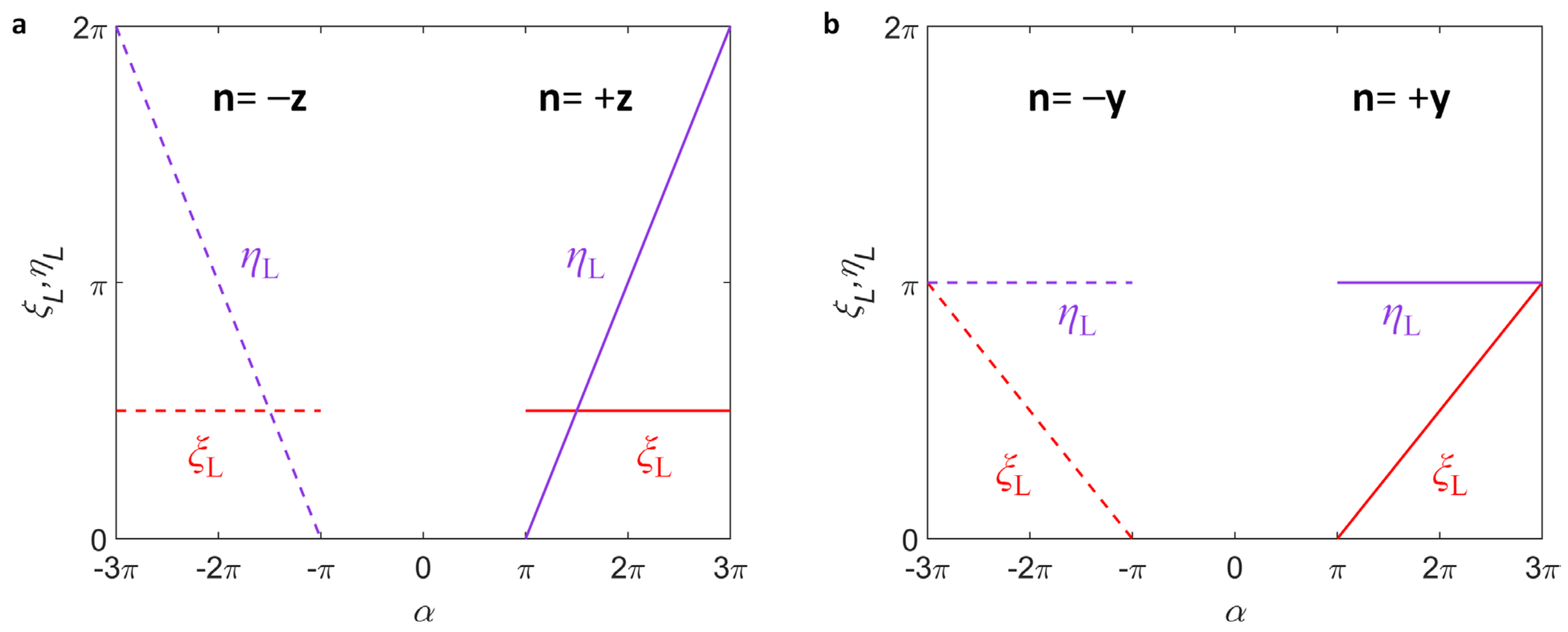

**Fig. S2. System parameters for fundamental operation modes. a,b,** Necessary phase shift values $\xi_L$ and $\eta_L$ for the rotations about the $z$-axis (**c**) and $y$-axis (**d**). Solid and dashed lines represent the cases of the positive and negative $z$- and $y$-axes, respectively.

Notably, the building block supports other rotation axes on the Bloch sphere, though the rotation angles are constrained. The allowed rotations can be separated into three cases: **n** on the $xy$-plane, $zx$-plane, and all octants except for the $yz$ plane. In discussing these cases, we represent **n** using spherical coordinates ($\theta_{\mathbf{n}}$, $\varphi_{\mathbf{n}}$) as follows: $\mathbf{n} = \mathbf{x}\sin\theta_{\mathbf{n}}\cos\varphi_{\mathbf{n}} + \mathbf{y}\sin\theta_{\mathbf{n}}\sin\varphi_{\mathbf{n}} + \mathbf{z}\cos\theta_{\mathbf{n}}$.

First, the rotation axis on the $xy$-plane is defined by setting $\theta_{\mathbf{n}} = \pi/2$, which leads to

$$U_{mn} = \sigma_0 \cos\frac{\alpha}{2} - i\sigma_x \cos\varphi_{\mathbf{n}} \sin\frac{\alpha}{2} - i\sigma_y \sin\varphi_{\mathbf{n}} \sin\frac{\alpha}{2}, \tag{S7}$$

from Eq. (4) in the main text. Because Eq. (S7) satisfies Eq. (5) in the main text, two conditions are necessary: $\xi_{\mathrm{L}} = l_\xi \pi$ and $\alpha = (2l_\alpha + 1)\pi$, where $l_\xi$ and $l_\alpha$ are integers. To realize all possible unitary operations about $\mathbf{n}$ on the $xy$-plane, it is sufficient to set $l_\xi = 0$ and $l_\alpha = 0$, with the range of $\varphi_{\mathbf{n}}$ being $0 \leq \varphi_{\mathbf{n}} \leq \pi$. Consequently, the required local phase shifts, when selecting the branch for $\eta_{\mathrm{L}}$ to result in the smallest value of the phase shift, are given by:

$$\xi_{\mathrm{L}} = 0, \qquad \eta_{\mathrm{L}} = 2(\pi - \varphi_{\mathbf{n}}), \tag{S8}$$

for $\alpha = \pi$ rotations.

Similarly, the rotation axis on the $zx$-plane is set by $\varphi_{\mathbf{n}} = 0$ or $\varphi_{\mathbf{n}} = \pi$, leading to

$$U_{mn} = \sigma_0 \cos\frac{\alpha}{2} \mp i\sigma_x \sin\theta_{\mathbf{n}} \sin\frac{\alpha}{2} - i\sigma_z \cos\theta_{\mathbf{n}} \sin\frac{\alpha}{2}. \tag{S9}$$

In aligning Eq. (S7) with Eq. (5) in the main text, we can set $\alpha = \pi$ and $\varphi_{\mathbf{n}} = 0$. For $0 \leq \theta_{\mathbf{n}} \leq \pi$, the necessary local phase shifts are specified as:

$$\xi_{\mathrm{L}} = \theta_{\mathbf{n}} + \frac{\pi}{2}, \qquad \eta_{\mathrm{L}} = 0. \tag{S10}$$

In contrast, for the rotation axis on the $yz$-plane, which is defined by $\varphi_{\mathbf{n}} = \pi/2$ or $\varphi_{\mathbf{n}} = 3\pi/2$, Eq. (4) in the main text becomes

$$U_{mn} = \sigma_0 \cos\frac{\alpha}{2} \mp i\sigma_y \sin\theta_{\mathbf{n}} \sin\frac{\alpha}{2} - i\sigma_z \cos\theta_{\mathbf{n}} \sin\frac{\alpha}{2}. \tag{S11}$$

Notably, $\alpha = 2l_\alpha\pi$ except for $\mathbf{n} = \pm\mathbf{z}$ and $\mathbf{n} = \pm\mathbf{y}$ axes. Therefore, there are no possible rotation operations on the Bloch sphere for $\mathbf{n}$ on the $yz$-plane, apart from the $\mathbf{n} = \pm\mathbf{z}$ and $\mathbf{n} = \pm\mathbf{y}$ axes.

For the other octant regions except for the $xy$-, $yz$- and $zx$-planes, we calculate the allowed rotation angle $\alpha$ and the required local phase shifts $\xi_L$ and $\eta_L$, for a given $\mathbf{n}$, by comparing Eqs. (4) and (5) in the main text. In detail, local phase shifts are determined by $\mathbf{n}$, as follows:

$$\xi_L = \arctan(-\frac{\cos\theta_{\mathbf{n}}}{\sin\theta_{\mathbf{n}}\cos\varphi_{\mathbf{n}}}), \qquad \eta_L = 2\arctan(\tan(-\varphi_{\mathbf{n}})), \tag{S12}$$

in the ranges of $0 \le \theta_{\mathbf{n}} \le \pi$ and $0 \le \varphi_{\mathbf{n}} < 2\pi$ except for the *xy*-, *yz*- and *zx*-planes. With the obtained phase shifts, the allowed rotation angle is determined as follows:

$$\alpha = 2\arccos(-\sin\xi_L \sin\frac{\eta_L}{2}), \tag{S13}$$

in the range of $0 \le \alpha < 2\pi$. The results are illustrated in Fig. S3, which are also aligned with the results in Fig. S2.

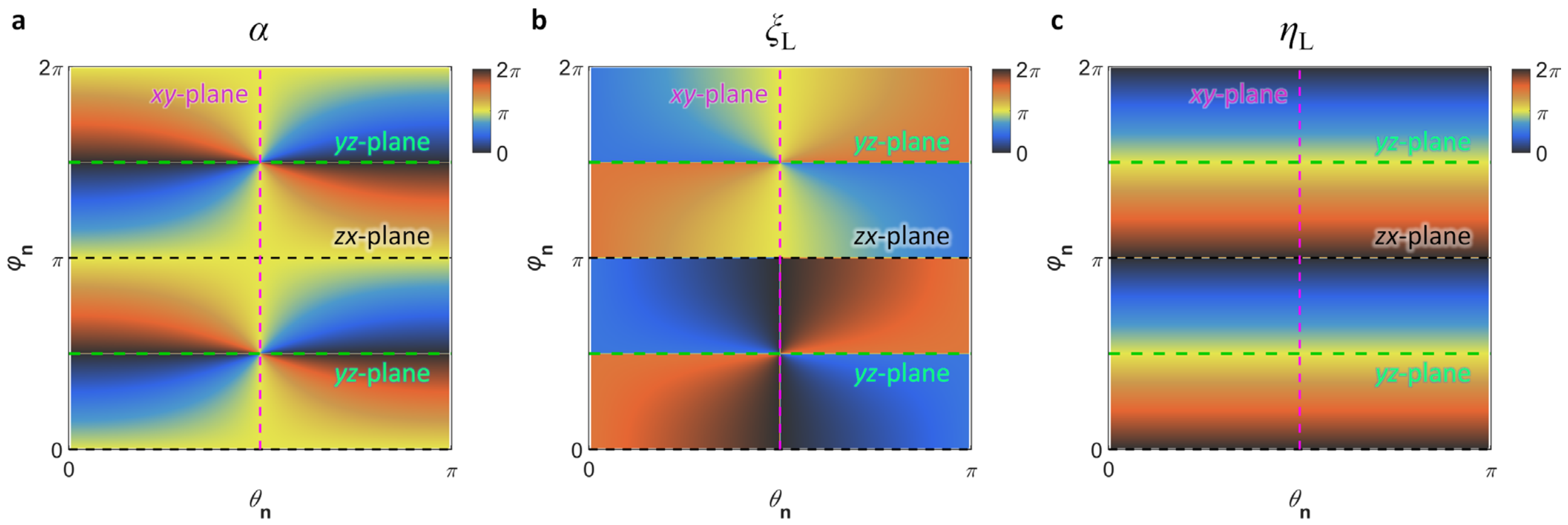

**Fig. S3. Accessible rotation operations and system parameters for given rotation axes. a,** Rotation angle $\alpha$. **b,c,** Necessary phase shift values $\xi_L$ (**b**) and $\eta_L$ (**c**). The angles and phase shifts are calculated for each $\mathbf{n}$, which is characterized by its spherical coordinates $\theta_{\mathbf{n}}$ and $\varphi_{\mathbf{n}}$. The green dashed lines denote the rotation axis $\mathbf{n}$ on the *yz*-plane, representing the prohibited cases in our building block, except for the *y*- and *z*-axes.

**Note S3. Numerical design of building blocks**

To demonstrate the experimental validity of our building block, we conduct a numerical analysis for its implementation in integrated photonics. We apply the three-dimensional (3D) finite-difference time-domain (FDTD) method and two-dimensional (2D) finite-difference frequency-domain (FDFD) method in the commercial software Tidy3D[9] to estimate the device footprint, coupling strength, operation bandwidth, and losses. Figure S4 illustrates the schematic of the overall structure for our building block. The lines represent silicon waveguides (relative permittivity 12.11) with a height of 0.22 μm and width of 0.50 μm on a $SiO_2$ substrate (relative permittivity 2.10), utilizing the fundamental transverse electric (TE) mode. The circumference of the ring resonator is 80 μm, while the lengths of the reciprocal phase shifter (red lines) and nonreciprocal waveguide (yellow lines) are 160 μm and 9.0 mm, respectively. A cascaded connection of a reciprocal phase shifter and a nonreciprocal waveguide jointly operates as a nonreciprocal phase shifter. The total footprint of the building block designed by the FDTD and FDFD methods is approximately 0.62 $mm^2$.

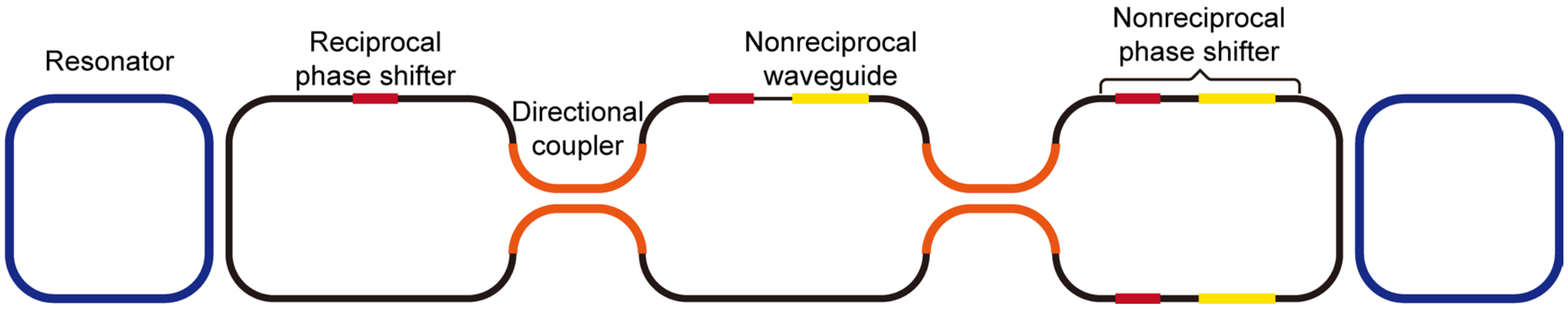

**Fig. S4. Building block for numerical design.** The figure presents a schematic of the entire structure of the building block. Red, yellow, and orange lines represent reciprocal phase shifters, nonreciprocal waveguides, and directional couplers, respectively. To freely manipulate NRPSs, a nonreciprocal phase shifter is realized through a combination of a cascaded reciprocal phase shifter and a nonreciprocal waveguide. Thermo-optical and magneto-optical modulations are applied to reciprocal and nonreciprocal phase shifts for reconfigurable phase shifts, respectively.

In the design of the proposed building block, four key components are included:

waveguide-resonator coupling, reciprocal phase shifters, nonreciprocal waveguides, and directional couplers. First, we analyse the coupling between the resonator and the waveguide segment of the loop coupler using the FDTD method to determine the decay rate $1/\tau$. In the analysis, we also examine the consequent radiation loss $\kappa_{\text{int}}$ that occurs through the coupling process. When developing the waveguide-resonator coupling setup, as depicted in Fig. S5a, for the FDTD simulation, the temporal coupled-mode theory modelling the setup leads to the following relationship between the input ($v^{+}$) and output ($v^{-}$) waves, as follows:

$$S_{21} \equiv \frac{v^{-}}{v^{+}} = Ae^{i\theta(\omega)} \frac{2\left[i(\omega-\omega_0)-\kappa_{\text{int}}\right]\tau+1}{2\left[i(\omega-\omega_0)-\kappa_{\text{int}}\right]\tau-1}, \tag{S14}$$

where $A$ and $\theta(\omega)$ represent the amplitude and frequency-dependent phase offsets, respectively, describing wave propagation through the input waveguide in the absence of the resonator, and $\omega_0$ denotes the resonance frequency. The model approximates the FDTD-calculated results near the resonance (Fig. S5b,c) with the parameters $A = -0.16$ dB, $1/\tau = 617 \times 10^{9}$ s$^{-1}$, and $\kappa_{\text{int}} = 17.6 \times 10^{9}$ s$^{-1}$. The numerical analysis indicates that the coupling strength between the resonator and the coupler waveguide is substantially greater than the radiation loss, which satisfies the necessary condition for the theory outlined in Note S1.

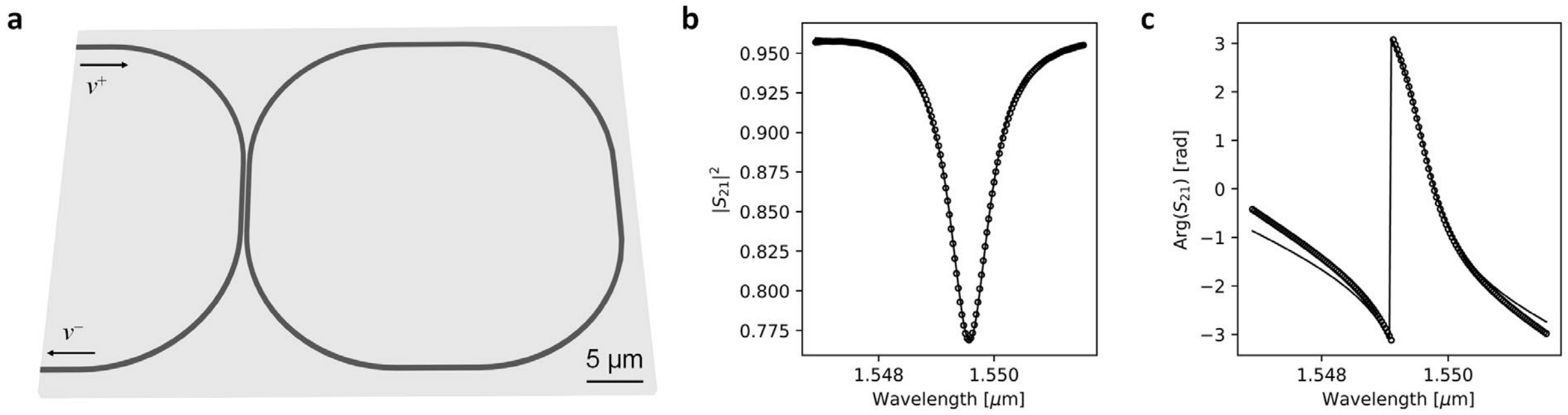

**Fig. S5. Waveguide-resonator coupling. a,** A schematic for the coupling analysis using the FDTD method. **b,c,** The intensity (**b**) and phase (**c**) of the scattering parameter $S_{21}$ between the input and output waves.

The next component is the reciprocal phase shifter, which employs the thermo-optical modulation of the silicon waveguide[10,11]. We set the range of relative permittivity modulation at ±0.50 %, an experimentally verified value[11,12]. We apply the FDFD mode solver to estimate changes in the propagation constant along the waveguide. The numerical results indicate that a modulation length of 16 μm is sufficient to achieve a $2\pi$ reconfigurable phase shift (Fig. S6a), while guaranteeing operation within the 2.5 THz bandwidth (Fig. S6b). Furthermore, by applying the FDTD method, we investigate the interface loss between the modulated and unmodulated regions at the maximal permittivity mismatch, which is approximately 0.05 dB.

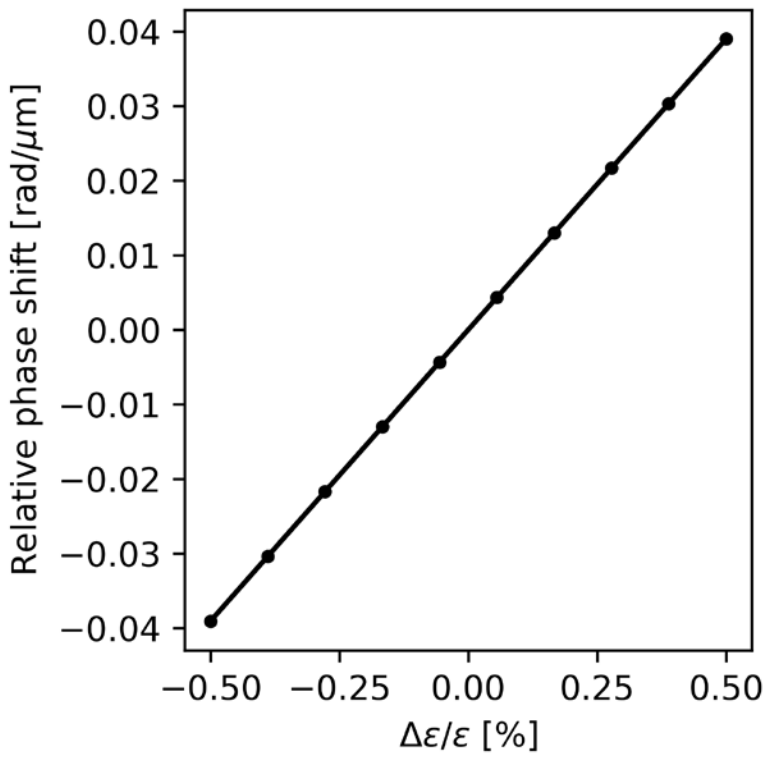

**Fig. S6. Reciprocal phase shifter. a,** The phase shift as a function of relative permittivity modulation, where $\Delta\varepsilon$ and $\varepsilon$ denote the modulation and the original value ($\varepsilon = 12.11$) of the relative permittivity of silicon, respectively. **b,** The phase shift as a function of the operating wavelength at $\Delta\varepsilon/\varepsilon = 0.50$ %.

We also utilize the FDFD method to evaluate the NRPS along the Ce:YIG-contacted silicon nonreciprocal waveguide. For the waveguide cross-section shown in Fig. 1e in the main text, the FDFD method leads to the propagation constants of the forward and backward propagating waves, which are tunable with the external static magnetic field and the following Faraday rotation. The results indicate that a waveguide length of 9.0 mm is adequate for achieving a $4\pi$ NRPS. Additionally, we also apply the FDTD simulation to estimate losses, revealing an interface loss of 0.33 dB and a propagation loss of 0.06 dB near the wavelength of 1550 nm.

The final component is the directional coupler used to form a Mach-Zehnder interferometer. The coupling length and the gap between the waveguides are 3.6 μm and 0.1 μm, respectively. To balance the loss and footprint of the coupler, we adopt a bending radius of 108 μm and a bending angle of $\pi/8$ for the directional coupler. The FDTD result shown in Fig. S7 yields a symmetrical transfer matrix at a 1550 nm wavelength with a 0.10 dB insertion loss.

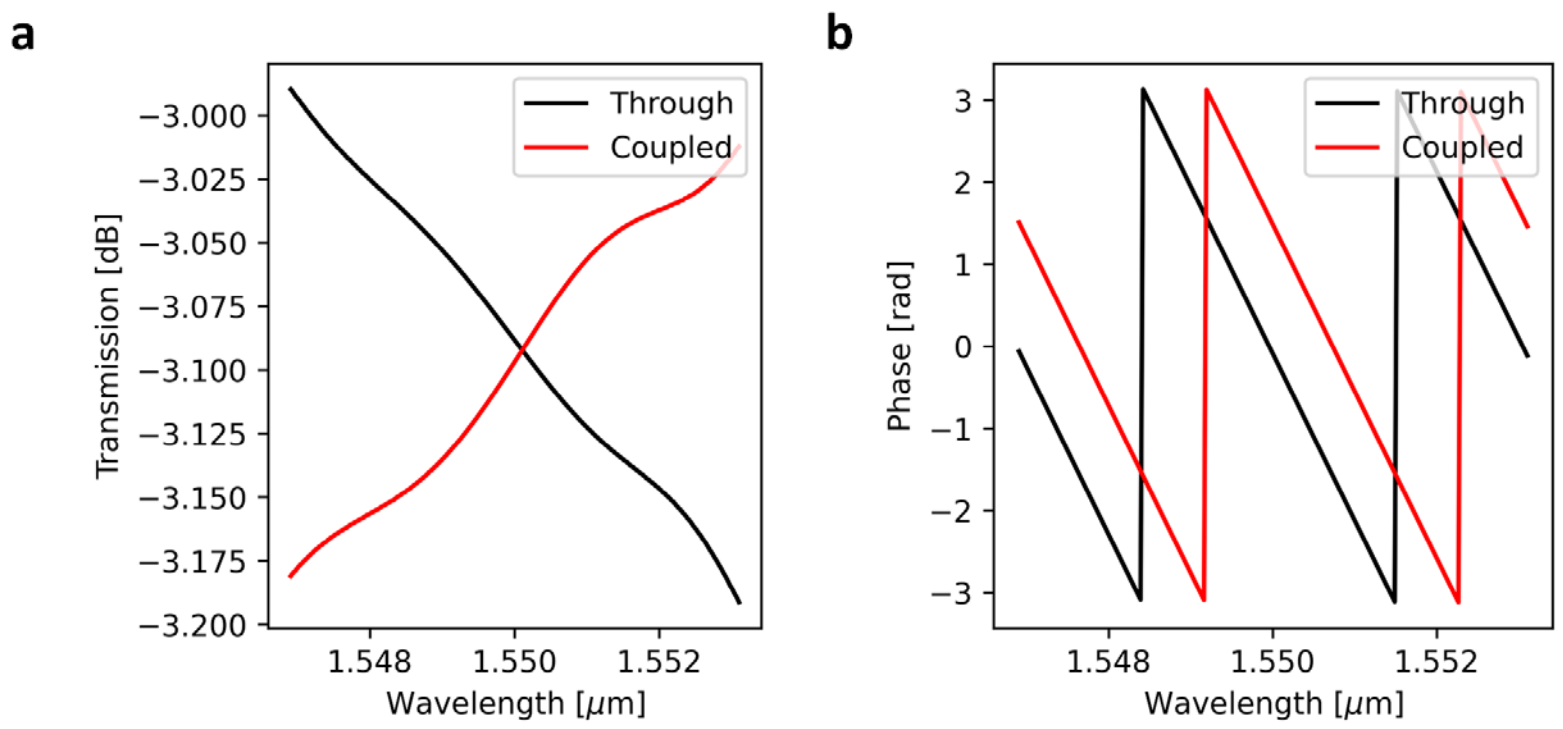

**Fig. S7. Directional coupler. a,b,** Directional coupler performance: transmission (**a**) and phase (**b**) calculated with the FDTD method around a 1550 nm wavelength. Black and red lines denote the through and forward coupled waves, respectively.

Based on the component designs listed, we estimate the total loss of the nonreciprocal loop coupler. The accumulated losses along the coupler is approximately 1.35 dB, which is reasonable given the designed coupling strength of $1/2\tau = 308.5\times10^9$ $s^{-1}$.

**Note S4. Spinor distribution of edge states**

The uniqueness of the non-Abelian interface, compared to an Abelian one, is also evident in the spatial distribution of edge-state spinors. At the Abelian interfaces (Fig. S8a,b), the bulk lattices around the interface share a common spinor basis. In this configuration, the gauge fields are block-diagonalized on each spinor, which results in a spatially homogeneous distribution of the spinor basis for each edge state. In contrast, at the non-Abelian interface (Fig. S8c), the Hamiltonian no longer adopts a common spinor basis. Therefore, the hybridized edge states, which contain spatially nonuniform contributions from the edge states of each spinor basis, display a spatially inhomogeneous spinor. This distinctive mixing of spinors is a hallmark of non-Abelian interfaces.

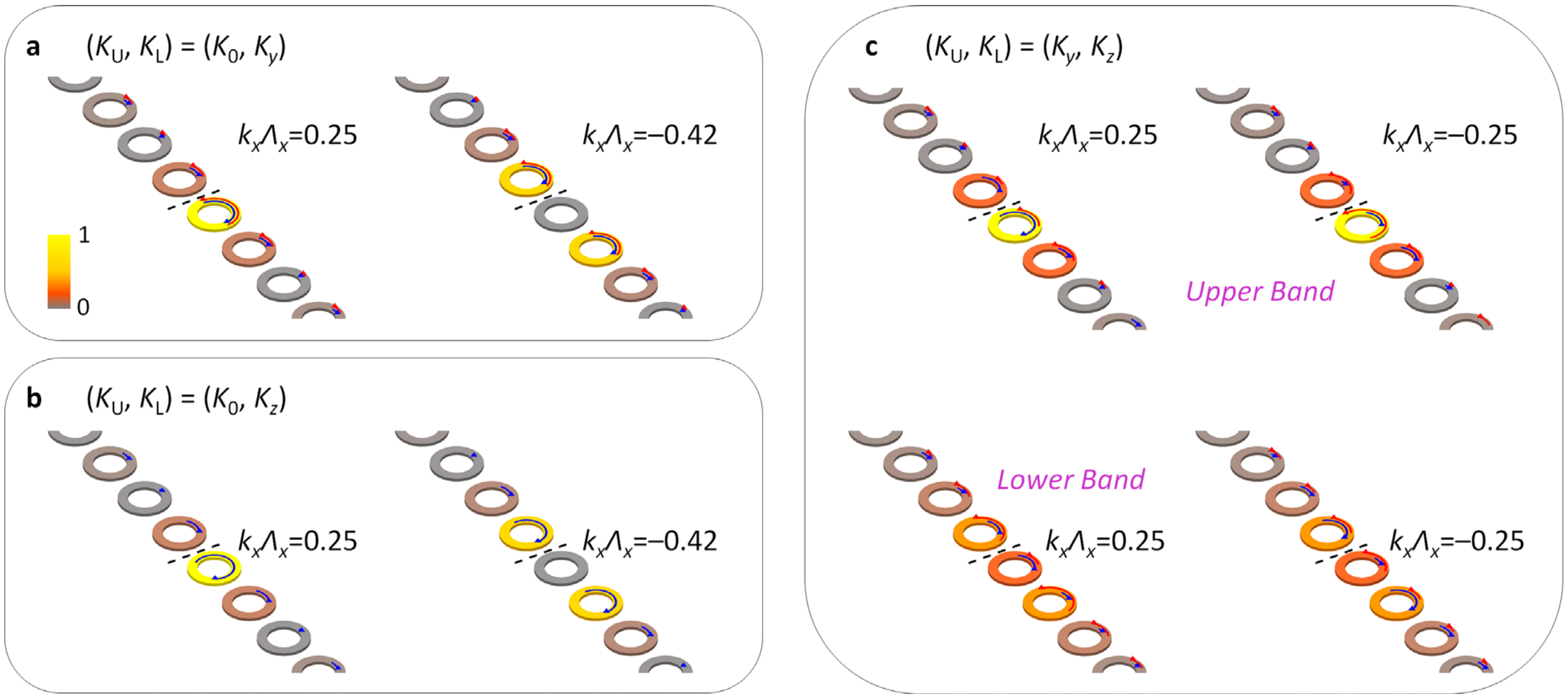

**Fig. S8. Abelian and non-Abelian edge states. a-c,** Edge state profiles for the Abelian cases ($K_0$, $K_y$) (**a**) and ($K_0$, $K_z$) (**b**), and the non-Abelian case ($K_y$, $K_z$) (**c**). In the Abelian cases, the ratios of amplitudes and phases between pseudospins, which are described by the lengths and locations of the red and blue arrows, respectively, are consistent across all resonators. In contrast, these ratios vary spatially in the non-Abelian case. The colour of each resonator illustrates the optical intensity of an eigenspinor at the resonator. Each edge state corresponds to a black empty circle depicted in the edge states of Fig. 3e-g in the main text. Black solid arrows in **c** describe the hybridization between topologically nontrivial edge states.

**Supplementary References**